\documentclass[acmsmall, screen]{acmart}
\AtBeginDocument{%
  }

\usepackage{amsmath,amssymb,amsfonts}
\usepackage{algorithmic}
\usepackage{graphicx}
\usepackage{textcomp}
\usepackage{cleveref}
\usepackage{booktabs}

\usepackage[table]{xcolor}
\usepackage{tabularx}
\usepackage{multirow} 

\usepackage{amssymb}    

\usepackage{listings}
\usepackage{tikz}
\usepackage{listings}
\usepackage[table]{xcolor}
\usepackage{tcolorbox}
\tcbuselibrary{listings}
\usetikzlibrary{shapes, arrows.meta, positioning}
\usepackage{amsmath}
\usepackage{makecell}
\usepackage{enumitem}

\usepackage{url}
\usepackage{multirow}

\usepackage{xspace}
\usepackage{booktabs}
\usepackage{tabularx}   
\usepackage{caption}
\usepackage{tabularray}
\usepackage{siunitx}

\usepackage[commandnameprefix=ifneeded,authormarkup=none]{changes}
\definechangesauthor[name={dan},color=red]{DN}

\begin{document}
\newcommand{\tool}{\textit{Path2Spec}\xspace}
\title{\tool: Path-Aware Specification Generation via Large Language Models\\
}
\author{Dan Huang}
\affiliation{%
  \institution{Singapore Management University}
  \country{Singapore}
}
\email{dan.huang.2024@phdcs.smu.edu.sg}

\author{Zhensu Sun}
\affiliation{%
  \institution{Singapore Management University}
  \country{Singapore}
}
\email{zssun@smu.edu.sg}

\author{Huihui Huang}
\affiliation{%
  \institution{Singapore Management University}
  \country{Singapore}
}
\email{hhhuang@smu.edu.sg}

\author{Jinfeng Jiang}
\affiliation{%
  \institution{Singapore Management University}
  \country{Singapore}
}
\email{jfjiang@smu.edu.sg}

\author{Xiaofei Xie}
\affiliation{%
  \institution{Singapore Management University}
  \country{Singapore}
}
\email{xfxie@smu.edu.sg}

\author{Yintong Huo}
\affiliation{%
  \institution{Singapore Management University}
  \country{Singapore}
}
\email{ythuo@smu.edu.sg}

\author{David Lo}
\affiliation{%
  \institution{Singapore Management University}
  \country{Singapore}
}
\email{davidlo@smu.edu.sg}


\fancyfoot{} 
\renewcommand{\shortauthors}{Huang et al.}

\pagenumbering{arabic} 

\begin{abstract}
Formal specifications are critical for program verification, comprehension, and maintenance. However, manually writing them is costly and difficult to scale. Recent studies have shown that Large Language Models (LLMs) are
promising for automated specification generation, but existing methods suffer from quality issues. We analyze a state-of-the-art approach and find that at least 34.6\% of successfully verified specifications actually fail to meaningfully capture the program’s distinct behavior, which is a quality issue not captured by metrics that only measure verification success. 
We further found that a major factor contributing to such hidden quality issues stems from the design of existing methods: these methods treat a program as a single unit, resulting in overly general, coarse-grained constraints.
To this end, we introduce \tool, a divide-and-conquer framework that addresses these
limitations through systematic path-based reasoning. \tool leverages LLMs to extract all execution paths
from an input program, generates path-specific specifications for each, and merges them into a comprehensive
overall specification. For complex programs where path-based generation struggles, \tool employs a
decompose-then-retry strategy that recursively breaks a program into smaller subprograms based on logical
branches, generates specifications for each, and merges them back. We evaluate \tool on two public
benchmarks: SG-Bench (120 programs) and SV-COMP (265 programs).
Results show that~\tool can outperform the state-of-the-art baseline SpecGen: 87.5\% versus 66.7\% on SG-Bench, and 83.0\% versus 44.2\% on
SV-COMP.
Human evaluation further validates that \tool generates higher-quality specifications with precise semantic alignment to the code. Our work demonstrates that our path-based specification generation strategy
substantially enhances the quality and accuracy of LLM-generated specifications.
\end{abstract}

%%
%% The code below is generated by the  at http://dl.acm.org/ccs.cfm.




\title{
\tool: Path-Aware Specification Generation via Large Language Models
}
\maketitle
\section{Introduction}
\label{sec:intro}
Formal specifications are machine-readable descriptions of a program’s expected behavior and intent. Common specifications, such as preconditions (function input requirements), postconditions (output properties), and loop invariants (conditions preserved across iterations), serve as concise annotations to source code. Such specifications are fundamental to program comprehension, formal verification, and long-term maintenance. For instance, preconditions and postconditions can help automatically generate unit tests for code~\cite{dakhel2024effective}.

Writing formal specifications for a specific program is challenging because it relies heavily on human experts on human experts, which is both expensive and hard-to-scale. To this end, automated specification generation tools have been proposed.
For example, Daikon~\cite{ernst2007daikon} uses dynamic analysis to infer likely invariants by statistically generalizing over large execution traces, whereas Houdini~\cite{flanagan2001houdini} applies static, theorem-prover-driven inference that iteratively eliminates counterexamples. 
More recently, researchers have explored using Large Language Models (LLMs) for specification synthesis~\cite{ge2023openagi}. For example, AutoSpec~\cite{autospec} utilizes LLMs to generate candidate specifications, which are subsequently verified through program verification to filter out incorrect ones. As an improvement over AutoSpec, SpecGen~\cite{specgen} employs a conversational approach for iterative enhancement and applies mutation operators with a heuristic selection strategy to refine and verify the specifications when the LLM fails.
Compared to traditional tools, these LLM-based methods have demonstrated superior performance and wider applicability, indicating a promising technical pathway.

\begin{figure*}[t]
\centerline{\includegraphics[width=0.9\linewidth]{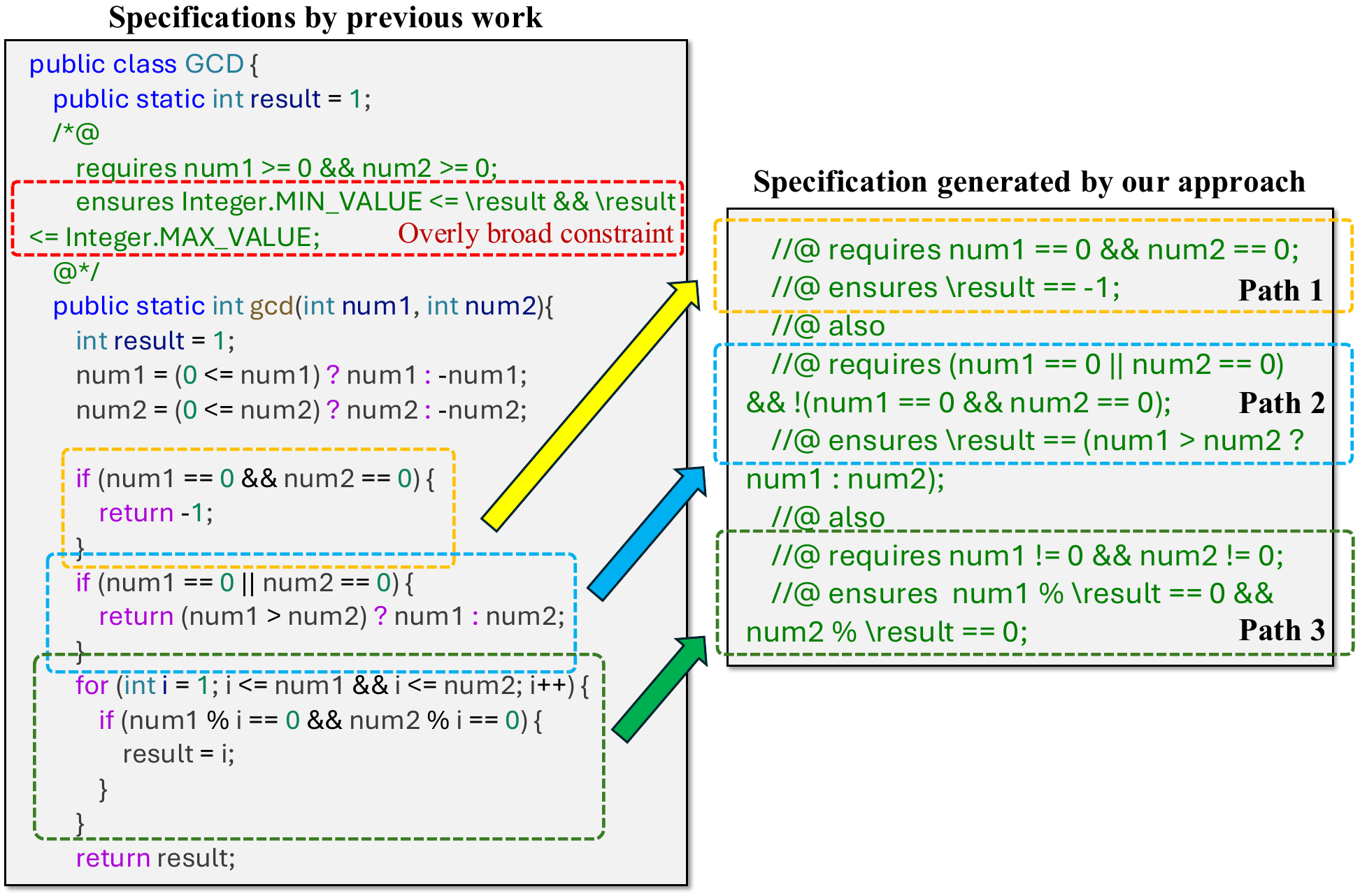}}
\caption{An example illustrating the problems in formal specifications generated by previous work, and demonstrates the more precise preconditions and postconditions obtained through extracting possible execution paths from the program.}
\label{wrong}
\end{figure*}

However, such LLM-based methods focus on generating valid specifications while ignoring their quality, especially in the coverage of the program's behaviors. Coverage refers to whether the formal specification provides corresponding descriptions for all possible execution scenarios of the program. If a program has many execution paths, but only a few of them are correctly described in a specification, we consider such a specification to have low coverage. 
This lack of coverage cannot be detected by standard verifiers. 
Verifiers merely determine if the postconditions are satisfied given the preconditions; this check does not guarantee that all possible execution behaviors have been captured.
To fill this gap, we propose a metric called Logical Strength.
It compares the relative logical quality of two (precondition, postcondition) pairs by evaluating the strictness of their constraints. The comparison focuses on whether the generated specification narrows the input range through its precondition and broadens the output range through its postcondition. Relative to the ground-truth specification, if the generated precondition admits a narrower range of inputs while the generated postcondition permits a broader range of outputs, then the generated (precondition, postcondition) pair is considered a logically incomplete specification.

Using this metric, we analyze specifications generated by the state-of-the-art tool SpecGen~\cite{specgen} against human-written ground-truth specifications from the SpecEval~\cite{ma2024speceval} benchmark, which contains 125 verified specifications corresponding to 173 (precondition, postcondition) pairs.
The analysis finds that SpecGen generates 107 (precondition, postcondition) pairs that pass verification. However, among these verified pairs, 34.6\% (37 pairs) are logically incomplete compared to ground truth, and 27.1\% (29 pairs) are undetermined(i.e., one component, such as the precondition, was stronger, while the other was weaker, indicating a trade-off rather than an improvement).
These findings empirically confirm that current methods frequently produce low-coverage specifications.

Moreover, we manually examine the generated specifications, which further reveals the main failure mode of this issue.
We illustrate this using an example.
In Figure~\ref{wrong}, the specification of the output is~\texttt{Integer.MIN\_VALUE <= r
esult \&\& result <= Integer.MAX\_VALUE}, a valid but overly broad postcondition since it is always true for all possible values.
However, the program has three distinct execution paths with each satisfying its unique constraints, i.e., three different (precondition, postcondition) pairs in total. Therefore, although the LLM generated a valid specification, it fails to capture the distinct execution paths of the program, thereby providing no meaningful constraint or insight into the program's true functionality.

This limitation originates from the design of existing approaches such as SpecGen~\cite{specgen} and AutoSpec~\cite{autospec}, which typically treat the entire program as a single unit, prompting the model to generate specifications for the whole code at once. Without explicit path-level guidance, LLMs tend to produce coarse-grained specifications, i.e., overly broad method-level constraints that summarize the program as a whole. Although such specifications can easily pass the verifier's check, they often fail to distinguish different input regions and execution paths, and therefore cannot precisely characterize the program's path-specific outputs.

Therefore, although LLMs already demonstrate strong capabilities in understanding code semantics, generating specifications that precisely capture the intended program behavior remains a challenging and unsolved problem.

In this paper, we aim to address the challenges of generating both \textit{correct} and \textit{high-quality} program specifications. Specifically, we propose \tool, a \textit{divide-and-conquer} framework designed to generate specifications that accurately capture fine-grained program behaviors. By systematically forcing an LLM to reason over individual execution paths, our approach ultimately produces path-specific and highly detailed specifications.
Specifically, \tool first uses an LLM to extract all distinct execution paths from the input program.
It then generates a path-specific specification for each, which is subsequently merged to construct the overall program specification.
This path-level reasoning significantly improves the precision and detail of the output.
However, when the program is too complex, LLMs may struggle to generate correct specifications, even when guided by a specific path.
Therefore, we further propose a \textit{decompose-then-retry} strategy.
For a program that an LLM fails to produce correct path-based specifications, we decompose it into smaller subprograms based on its logical branches and retry the path-based generation to produce specifications for each subprogram.
This decompose-then-retry process repeats if the generation for any subprogram still fails.
Finally, the specifications for all subprograms are merged to form the overall specifications for the original input program.

To evaluate the usefulness of \tool, we have conducted extensive experiments on two public datasets used by prior works, SpecGenBench (SG-Bench)~\cite{specgen} and SV-COMP~\cite{beyer2024state}, respectively comprising 120 and 265 Java programs. 
% We largely follow the setting of the state-of-the-art baseline SpecGen~\cite{10.1109/ICSE55347.2025.00129} and focus on Java.
The experimental results demonstrate that \tool achieves substantial improvements. 
On SG-Bench, \tool achieves 87.5\% pass@1 compared to the baseline's 66.7\%. On SV-COMP, \tool reaches 83.0\% versus SpecGen's 44.2\%.
Additionally, we perform an ablation study to separately validate the effectiveness of providing path information as guidance (Path-only) and decomposing programs to reduce program complexity (Decomposition-only) across the same four base LLMs on both benchmark datasets.
The results confirm that both Path-only and Decomposition-only strategies outperform the baseline across both datasets, demonstrating the positive impact of each component. On SG-Bench, Path-only achieves 82.5\% and Decomposition-only achieves 77.5\%, while on SV-COMP, Path-only reaches 81.1\% and Decomposition-only achieves 60.8\%. Notably, neither strategy alone surpasses \tool's performance. Therefore, combining both methods further strengthens \tool's capability in generating formal specifications. Furthermore, human evaluation shows that \tool generates higher-quality specifications than the baseline, achieving an average score of 3.4 versus 2.4 (out of 4). The generated specifications are semantically aligned with programs and exhibit more comprehensive coverage of actual program execution behaviors.

Our main contributions can be summarized as follows:
\begin{itemize}[leftmargin=*]
    \item We quantitatively and qualitatively analyze the quality issues of existing LLM-generated specifications on the SpecEval benchmark. We show that 34.6\% of specifications that pass verification are actually incomplete compared to their human-written counterparts, a weakness obscured by existing evaluation metrics.
    \item To fill this gap of the low-coverage issue, we propose \tool, a divide-and-conquer framework for generating fine-grained, high-quality program specifications.
    \item To demonstrate the effectiveness of \tool, we conduct extensive experiments on two benchmarks (SG-Bench and SV-COMP) with multiple base LLMs, demonstrating substantial performance improvements through path-guided generation. The overall performance across the two datasets improves from 51.2\% to 84.4\%, and human evaluation confirms that our generated formal specifications exhibit substantially higher quality.
\end{itemize}

The rest of the paper is organized as follows. Section 2 introduces the background and motivation of this study. Section 3 presents our systematic approach and illustrative examples. Section 4 describes our experimental setup, datasets, and evaluation metrics. Section 5 provides a comprehensive analysis of the experimental results. Section 6 discusses the limitations of this study and the challenges that need to be addressed in future work. Section 7 reviews related work and provides a brief description of each. Section 8 concludes this paper and outlines future research directions.
\section{Background and Motivation}
In this section, we briefly review automated specification generation methods and reveal the low-coverage issue of existing LLM-based methods.

\subsection{Background}
\label{sec:background}
Conventional formal specification generation relies on symbolic or rule-based techniques, which can be categorized into dynamic and static methods. Dynamic approaches (e.g., Daikon~\cite{ernst2007daikon}) infer likely specifications by analyzing execution traces. Their major advantage is that they do not require reasoning about the program's source-level semantics.
However, their heavy reliance on test-suite coverage often leads to false positives: an incomplete test suite may fail to exercise all program behaviors, yielding specifications that are too weak or that contain spurious invariants holding only for the observed executions rather than for all possible ones~\cite{ernst2007daikon}.
In contrast, static approaches (e.g., Houdini~\cite{flanagan2001houdini}) analyze source code, data flow, and control flow, and use theorem provers to verify candidate specifications instantiated from templates. This paradigm aims to be sound for all possible inputs and thus offers stronger correctness guarantees. Nevertheless, it is constrained by the expressiveness of its templates and the reasoning capability of the verifier, frequently resulting in false negatives and requiring manually provided auxiliary invariants. Moreover, both dynamic and static methods depend heavily on predefined syntactic templates (e.g., \texttt{x < y}, \texttt{arr.length > 0}). This fundamental limitation confines them to inferring simple invariants of specific forms, making them inadequate for discovering complex or domain-specific logical constraints.

Recent years have witnessed a rapid shift toward LLM-based techniques.
Having learned semantic patterns from vast code corpora, LLMs can directly generate flexible, human-readable specifications, e.g., in the Java Modeling Language (JML), without relying on predefined templates or fixed grammars~\cite{bairi2024codeplan}. Existing work has progressively refined this paradigm. AutoSpec~\cite{autospec} adopts a generate-and-verify loop: the LLM proposes candidate specifications, and a verifier then filters out those that fail verification; this cycle repeats until all remaining specifications pass the verifier. SpecGen~\cite{specgen} goes one step further by feeding the error messages produced by the verifier (e.g., OpenJML~\cite{cok2011openjml}) back to the LLM, allowing it to refine its output incrementally and converge on verifiable annotations. Both approaches confirm that LLMs can move beyond fixed templates and rapidly synthesize sophisticated, machine-checkable specifications.
However, these methods focus solely on correctness, which does not guarantee the quality of the generated specifications. We elaborate on this limitation in the next subsection.

\subsection{Motivation}
\label{sec:motivation}
To reveal the quality issues of existing methods, we quantitatively and qualitatively analyze the performance of SpecGen on SpecEval~\cite{ma2024speceval}, a benchmark of 125 Java programs paired with verified human-written specifications, which together contain 173 (precondition, postcondition) pairs. Specifically, we run SpecGen with GPT-3.5-turbo as the base LLM and compare the generated specifications with their human-written counterparts.
This comparison follows a simple principle: relative to the ground truth, a generated specification should neither narrow the admissible input range (i.e., have a stronger precondition) nor widen the permitted output range (i.e., have a weaker postcondition). A stronger precondition admits fewer inputs and thus covers fewer program behaviors, whereas a weaker postcondition characterizes the expected outputs less precisely.

Accordingly, for each human-written (precondition, postcondition) pair, we check whether the LLM produces a corresponding pair whose precondition is at least as weak and whose postcondition is at least as strong, using an SMT-based implication check.
For example, if the human-written precondition and postcondition are \texttt{x >= 0} and \texttt{y < 0}, respectively, a generated pair of \texttt{x > 0} and \texttt{y <= 0} is considered \textit{incomplete}, since it fails to cover the input \texttt{x == 0} and additionally admits the output \texttt{y == 0}.
If the two components point in opposite directions (e.g., both the precondition and the postcondition are weaker, or both are stronger), we mark the pair as \textit{undetermined}.

Of the 173 pairs, 107 SpecGen-generated pairs pass verification. Among these verified pairs, 34.6\% are logically incomplete compared with the ground truth, and 27.1\% are undetermined, i.e., they exhibit trade-offs such as a weaker precondition paired with a weaker postcondition.
We therefore argue that current LLM-generated specifications suffer from significant quality issues, which are obscured by the existing evaluation metric (i.e., whether verification passes).

We further manually examine these LLM-generated specifications and observe one main failure mode: \textit{oversimplified specifications}. This occurs when the model fails to differentiate between distinct logical branches of the code and consequently merges branch-specific preconditions or postconditions into a single, overly general constraint.
As shown on the left of Figure~\ref{wrong}, the generated postcondition is \texttt{Integer.MIN\_VALUE <= result \&\& result <= Integer.MAX\_VALUE}, which holds for every \texttt{int} value and is thus trivially true.
However, the implementation involves three distinct execution paths, whose postconditions are \texttt{result == -1}, \texttt{result == (num1 > num2 ? num1 : num2)}, and \texttt{num1 \% result == 0 \&\& num2 \% result == 0}, respectively.
Therefore, although the generated postcondition is valid, it is insufficient, since each path has its own specific constraint on the output. A single oversimplified constraint over all inputs cannot meaningfully capture the semantics of the program.

These issues stem from the design of existing LLM-based specification generation methods. On the one hand, LLMs are instructed to process the code as a whole without path-specific guidance, and thus overlook path-specific constraints. On the other hand, for complex programs on which LLMs fail to generate correct specifications, they tend to weaken or drop constraints until the verifier accepts them. Consequently, the resulting specifications either fail to capture the program's actual behavior or impose excessively general constraints.

Our observations suggest that this issue is inherent to the design rather than to a particular base model or verifier. Re-running the official SpecGen artifact under the same OpenJML/CVC4 configuration as ours, we find that a large fraction of its \emph{verified} specifications are vacuous: they contain no postcondition or only \texttt{ensures true}, and pass verification merely because there is nothing to check. This fraction ranges from 25\% to 33\% with GPT-3.5-turbo and from 13\% to 19\% with DeepSeek-R1 on the two benchmarks used in our evaluation. For instance, SpecGen passes verification on the branching program \texttt{if\_expr1} with an empty specification, exactly where per-branch postconditions are required. In contrast, the single-shot variant of \tool, which localizes generation to individual paths, produces no vacuous verified specifications (0\%).

Generating specifications path by path makes the task substantially clearer. As shown in Figure~\ref{wrong}, three execution paths can be extracted from the source code, and a (precondition, postcondition) pair can be defined for each of them. In this way, the specifications are grounded in case-based analysis, improving their coverage of program behaviors. Motivated by this observation, we propose \tool, a path-based divide-and-conquer framework for LLM-powered specification generation.

\section{APPROACH}
\begin{figure}[htbp]  
  \centering
  \includegraphics[width=0.85\linewidth]{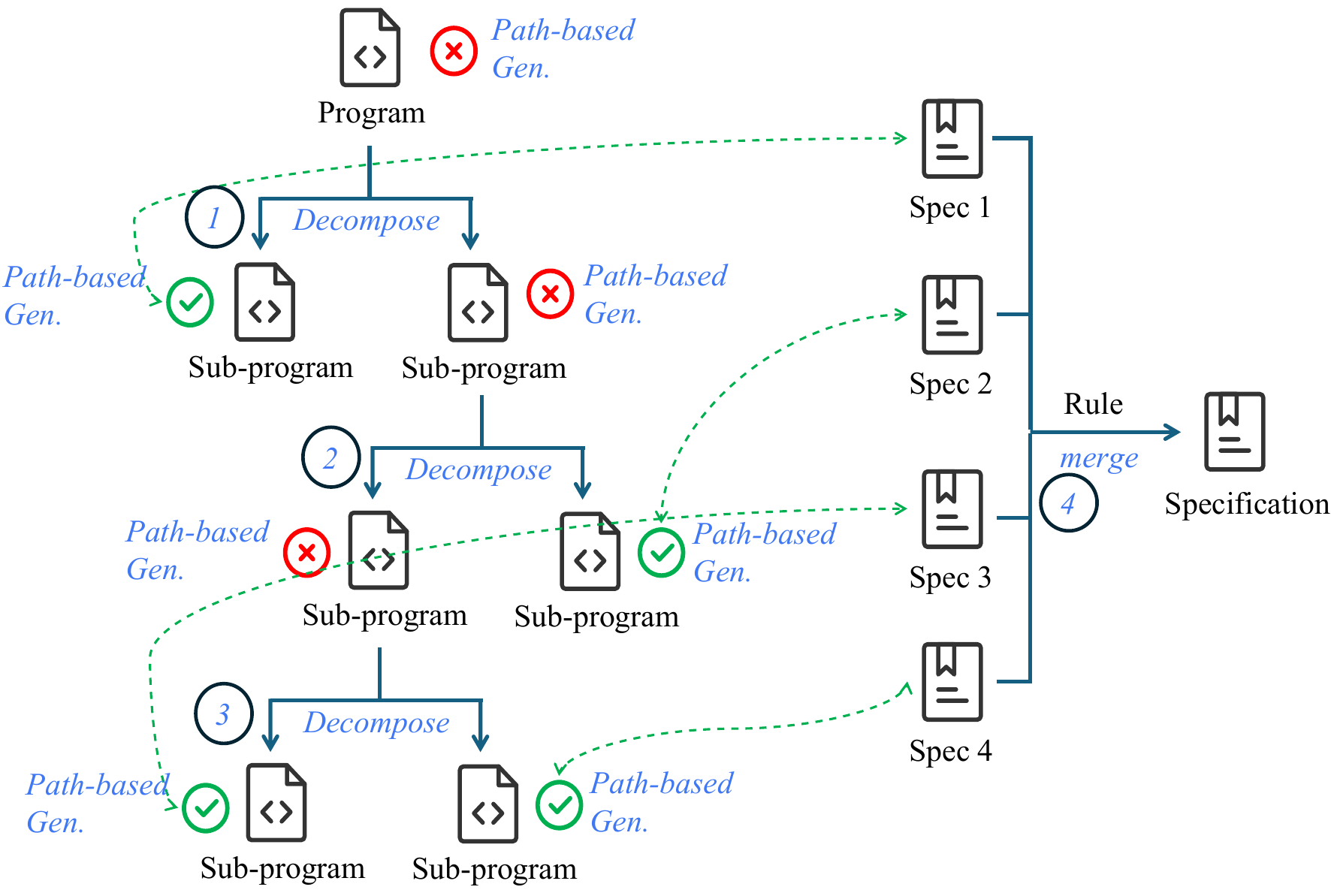} 
  \caption{Workflow of \tool. A red cross indicates that the path-based specification generation process failed, whereas a green check indicates success, which yields a set of specifications. When a program fails to generate a verified specification, it enters the first decomposition process (marked as \textcircled{1}). The program is decomposed into subprograms based on the first logical branch encountered. Each subprogram then enters the path-based specification generation process. If a subprogram successfully generates a verified formal specification, that specification is recorded. However, if a subprogram still produces an incorrect formal specification, it undergoes another decomposition process (marked as \textcircled{2}). The resulting subprograms repeat the path-based specification generation process. If any subprogram continues to fail in generating a verified formal specification, it is further decomposed (marked as \textcircled{3}). This iterative process continues until all subprograms produce verified formal specifications. Finally, we progressively merge and verify the formal specifications from all subprograms layer by layer, ultimately integrating them into the original program (marked as \textcircled{4}).}
  \label{fig: overall}
\end{figure}

\tool is an LLM-driven specification generation framework that guides the generation process using the execution paths of the input program. It employs a path-based specification generation workflow paired with a decompose-then-retry strategy, as illustrated in Figure~\ref{fig: overall} (detailed in Sections~\ref{sec:pathgen} and~\ref{sec:decomposenretry}, respectively).

Specifically, given an input program, $\text{\tool}$ first applies the path-based specification generation workflow by instructing an LLM to extract all execution paths from the program and then generate specifications for each path. These path-specific specifications are then verified with the OpenJML verifier. If any specification fails verification, its verification results are sent back to the LLM to iteratively refine the failed specification. Only after all specifications successfully pass the verification process, the workflow is considered complete, and the specifications are merged into a final specification and returned as the output. However, if verification still fails after $\text{M}$ rounds of iterative refinement, we consider the program to be too complex for the LLM to perform the path-based specification generation.

Therefore, for such failed cases, we propose a decompose-then-retry strategy to reduce the program's complexity. Specifically, for a program that fails to yield correct specifications with the path-based specification generation, we decompose it into two smaller subprograms based on its first logical branch. We then retry the aforementioned path-based specification generation workflow to produce corresponding specifications for each subprogram. The generated specifications for these subprograms are independently verified. If all subprogram specifications are successfully verified, they are combined to form the final specification for the original program. Conversely, if any subprogram specification still fails verification after iterative refinement, that subprogram is further decomposed based on its first logical branch, and the generation workflow is repeated. This decompose-then-retry process repeats recursively until either all resulting subprograms are successfully verified and combined, or until a subprogram cannot be further decomposed but still fails verification, at which point the original program is considered unresolvable.

Finally, after completing the path-based specification generation, we merge all path-specific specifications together. We primarily employ an interleaved merge-and-verify approach: we first merge the specification from the first path and invoke the verifier. Upon successful verification, we proceed to merge the specification from the second path, continuing this process until all path-specific specifications are successfully merged into the original program. For complex programs, since we initially apply a decompose-then-retry strategy, the merging process is conducted in two steps: first, we progressively integrate all path-specific specifications within each subprogram; then, we progressively integrate the specifications from subprograms into the original program in a bottom-up manner. Likewise, after completing the merge at each layer, we employ the verifier for validation, continuing this process until all successfully verified specifications from subprograms are merged into the original program.

\subsection{Path-Based Specification Generation}\label{sec:pathgen}
\begin{figure}[htbp]  
  \centering
  \includegraphics[width=0.9\linewidth]{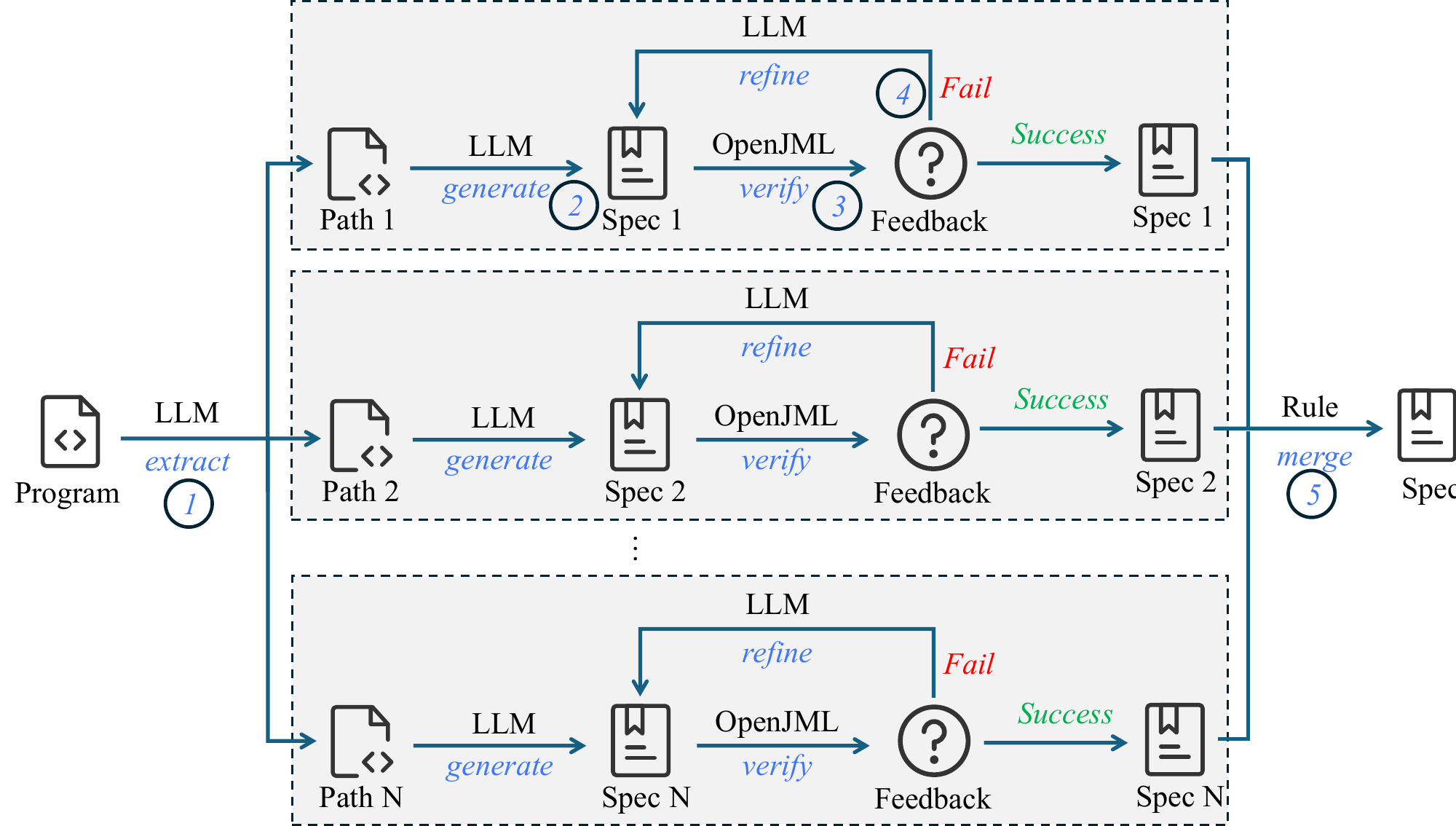}
  \caption{The workflow in the Path-based Specification Generation phase operates as follows: First, it extracts all possible execution paths from the program, which may contain ~\texttt{N} executable paths \textcircled{1}. Then, using the information from each execution path as guidance, it prompts the LLM to generate a formal specification for that path \textcircled{2}. The generated formal specification is verified using the OpenJML verifier \textcircled{3}. If verification fails, the error information is incorporated into the prompt to guide the LLM in refining the specification \textcircled{4}. This iterative refinement-and-verification process continues until correct specifications are generated for all paths, which are then merged into the original program \textcircled{5}.}
  \label{fig:pathgen}
\end{figure}

The workflow of the Path-Based Specification Generation Module is illustrated in Figure~\ref{fig:pathgen}.
It first extracts all possible execution paths from the input program to guide further generation.
The extraction is performed using an LLM.
Initially, we experimented with symbolic execution tools like Java PathFinder~\cite{mehlitz2013hands,puasuareanu2010symbolic} for more reliable path information. An execution path is a complete sequence of program statements from entry to exit, including all control flow decisions (conditionals, loops) taken during execution. Each path is characterized by its preconditions (input constraints that trigger the path) and postconditions (output expressions or states after execution). Therefore, the path information we aim to extract consists of (Precondition, Postcondition) pairs, where the precondition defines the input conditions under which the path is executed, and the postcondition specifies the resulting output expression.

However, current tools are built on dynamic execution engines and rely on analyzing dynamic execution data to infer results~\cite{10.1145/3650212.3680333, laursen2024reducing}. This approach has two key drawbacks: First, it depends on dynamic execution to infer results, and the effectiveness of dynamic execution is constrained by the path coverage of the test data. Second, the path analysis reports use concrete output values as postconditions rather than symbolic expressions. This type of path information provides limited benefit for LLM-based path-specific generation. Consequently, we adopt an LLM to analyze and extract execution paths, which yields more concise and interpretable path information that benefits the downstream workflow. We experimentally compare the performance of using an LLM versus Java PathFinder~\cite{mehlitz2013hands} for path extraction in~\Cref{sec:rq3}, providing empirical evidence to support our design choice.

Specifically, we prompt the LLM to generate all possible execution paths along with their corresponding preconditions (i.e., required input values) and postconditions (i.e., expected outcomes). The LLM adopts a symbolic execution approach, analyzing the program symbolically to derive corresponding output expressions for different input conditions. The main prompt used for this task is shown below:
\begin{tcolorbox}[size=title]
   \textbf{System Prompt (Path Extraction):} \\You are a symbolic execution analyst; your task is to analyze all
execution paths of the provided code using symbolic execution principles. Start from the method entry point, generate all possible paths based on Conditional statements, Loops, Branching constructs.\\
For each method: Generate ALL possible execution paths.\\
For each path:\\
Input: Clearly state the input condition (precondition) that triggers this path.\\
Output: Provide the output expression or state that results from executing this path with the given input condition.\\
Prohibited: Intermediate variables in the final output.\\
\textit{FINAL OUTPUT:}\\
\textit{Path [N]:}\\
\textit{Input: [condition]}\\
\textit{Output: [expression]}
\end{tcolorbox}

% \begin{table}[htbp]
% \centering
% \label{pu}
% \begin{tabularx}{\columnwidth}{|>{\columncolor{gray!20}}X|}
% \hline
% \\
% \normalsize \textbf{System Prompt (Path Extraction):} \\You are a symbolic execution analyst; your task is to analyze all
% execution paths of the provided code using symbolic execution principles. Start from method entry point, generate all possible paths based on Conditional statements,Loops, Branching constructs and so on.\\
% For each method: Generate ALL possible execution paths.\\
% For each path: Clearly state the input condition.\\
% Output expressions: Must be in terms of input variables only.\\
% Prohibited: Intermediate variables in final output.\\
% FINAL OUTPUT:\\
% Path [N]:\\
% Input: [condition]\\
% Output: [expression]
% \\
% \\
% \hline
% \end{tabularx}
% \end{table}
To reduce complexity and keep the LLM's context window focused on a single executable path, we adopt a path-by-path generation approach. For each obtained execution path, we prompt the LLM with the program and the specific path's information to generate the corresponding preconditions and postconditions, which represent the input and output for that execution path, as well as specifications for other components that comprise this path, such as loops. The main prompt is shown as follows:

% \vspace{-1in}
\begin{tcolorbox}[size=title]
\small
   \textbf{System Prompt (Formal Specification Generation):} \\You are a JML formal specification expert, generate JML
specifications for the given Java program based on path information. Path
information is the inference of different behavior cases of the provided program.\\
You can use the preconditions and postconditions from the path information as below:\\
Path input → requires clause as precondition\\
Path output → ensures clause as postcondition\\
Ensure specifications are complete for all execution paths
\\

\textit{Path information:}\\
\textit{Path [N]:}\\
\textit{Input: [condition]}\\
\textit{Output: [expression]}
\end{tcolorbox}

After that, we have a set of specifications for all the extracted paths.
These specifications are then respectively verified using OpenJML, where the specifications that fail the verification need to be refined.
The refinement is an iterative process.
Specifically, we incorporate the feedback from OpenJML, i.e., the error messages, into the prompt and ask the LLM to improve the corresponding failed specification. The main prompt for refinement is as below:
\begin{tcolorbox}[size=title]
   \textbf{System Prompt (Formal Specification Refinement):} \\You are an expert in JML formal specification. Analyze the provided error information and refine the JML annotations in the program accordingly. The following rules must be strictly observed to avoid critical issues:\\
1. Syntax Rules\\
2. Forbidden Patterns (instant rejection)\\
3. Timeout Prevention\\
4. Implicit OpenJML Checks\\
5. If the error message originates from a precondition or postcondition, use the provided path information to deduce whether the current precondition or postcondition is consistent with the actual program behavior along that path.\\
\textit{Error information: [error message]}\\
\textit{program: [program with JML spec]}\\
\textit{Path: [path information]}
\end{tcolorbox}
These refined specifications are then verified and will be refined again if they still fail.
This iterative process will end when the specifications are successfully verified or the predefined iteration limit (e.g., five iterations) is reached.
If the specifications for all paths are finally verified to be correct, we combine all the specifications and integrate them into the original program as the output.
However, if the full comprehensive specification fails verification, we consider the current program too complex for the LLM to understand, and apply a decompose-then-retry strategy to reduce the task complexity.

\subsection{Decompose-Then-Retry Strategy}\label{sec:decomposenretry}

The decompose-then-retry strategy is a top-down, recursive approach for handling failures in path-based specification generation.
When an input program fails this generation process, the strategy decomposes it into simpler subprograms based on its first-level branch condition.
Applicable branches include if-else statements, switch statements, and loops (e.g., for, while).
To reduce complexity, loops are handled specifically: they are decomposed into two subprograms.
One represents the case where the loop condition is satisfied (and the body executes), and the other represents the case where the condition is not met (and the loop is skipped).
\begin{figure}[htbp]
\centerline{\includegraphics[width=\linewidth]{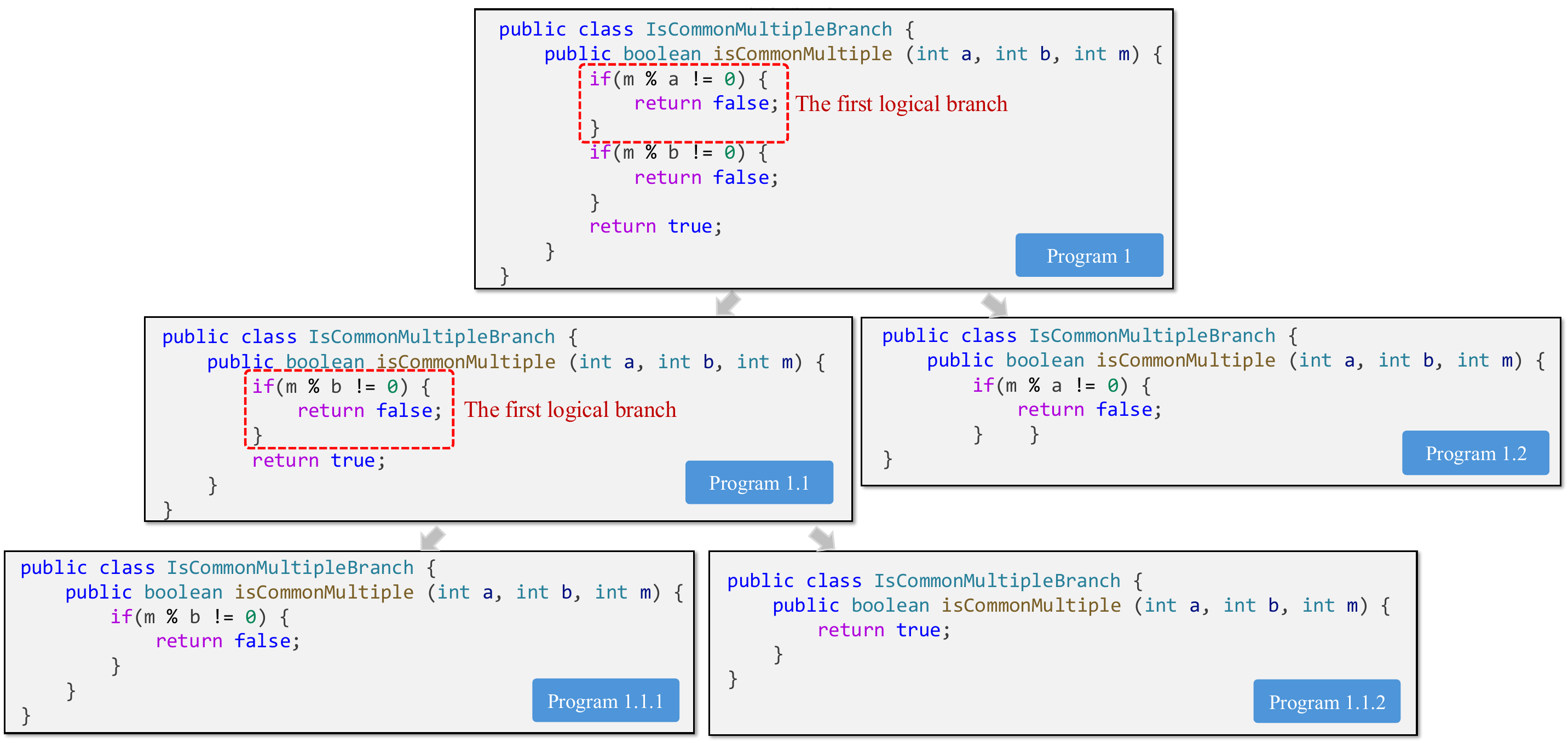}}
\caption{An example demonstrating how program~\texttt{1} is decomposed using a top-down strategy. The program is split based on its first logical branch into two subprograms (program~\texttt{1.1} and program~\texttt{1.2}). Each subprogram then undergoes path-based specification generation. If subprogram~\texttt{1.1} fails to generate correct formal specifications, it undergoes further decomposition at its first logical branch, creating smaller subprograms (program~\texttt{1.1.1} and program~\texttt{1.1.2}). This decompose-then-retry process repeats recursively until one of the following conditions is met: (1) all resulting subprograms successfully generate valid specifications, (2) all subprograms contain no control flow statements, or (3) the predefined iteration limit is reached.}
\label{fig:decompostion}
\end{figure}

Figure~\ref{fig:decompostion} illustrates this process. The initial input program (at the top of the figure) fails to generate a correct specification and is decomposed based on the \texttt{if(m \% a != 0) } branch into subprograms \textit{1.1} and \textit{1.2}.
The path-based generation is then retried on both resulting subprograms.
In this example, the resulting subprogram \textit{1.2} is simple, thereby easy to get valid specifications for an LLM.
Let's assume subprogram \textit{1.1} still fails.
For subprograms that fail in path-based specification generation, the strategy is applied again, further decomposing it based on its first branch, e.g., the \texttt{m \% b != 0} of this subprogram \textit{1.1}.
This decompose-then-retry process repeats recursively until one of the following conditions is met: all resulting subprograms are successfully verified, all subprograms contain no control flow statements, or the predefined iteration limit is reached.
In this example, the yielded subprograms \textit{1.1.1} and \textit{1.1.2} show significantly reduced complexity, lowering the barrier for the path-based generation to succeed.
However, if a subprogram cannot be further decomposed but still fails verification, we consider this case to be a failure and output existing valid specifications of other subprograms as a partial solution.
Compared with decomposing all possible branches at once, this top-down process can terminate early as soon as all subprograms at a certain depth are successful, saving unnecessary computational cost.

When the decompose-then-retry process ends, we collect all the valid path-specific specifications generated for each subprogram. We first integrate the verified path-specific specifications into their corresponding subprograms using an incremental merge-and-verify approach, where specifications are merged one by one with verification after each merge. Then, we integrate the specifications from subprograms into the original program using a bottom-up, layer-by-layer merge-and-verify approach. Then, we integrate the specifications from subprograms into the original program using a bottom-up, layer-by-layer merge-and-verify approach. 

Specifically, specifications from subprograms are merged into their respective parent programs one by one, with verification after each merge. This process proceeds upward until all specifications are integrated into the original program. During extraction, we record the contextual information for each specification, including which method the specification originates from, which loop statement it belongs to, and what the preceding and following lines of code are. We then classify the specifications into three main categories: Requires and Ensures (containing requires and ensures clauses), Loops (including loop invariants, maintaining, and decreases clauses), and Others (such as assignable, pure, signals, etc.). Finally, we insert them into the original program according to type-specific rules based on their category.
\begin{itemize}[leftmargin=*]
    \item \textbf{Requires and Ensures:} Requires and ensures from different methods are treated as different behaviors of the program and are concatenated using \texttt{//@ also}, then placed before the corresponding method with indentation aligned with the first line of code in that method.
    \item \textbf{Loops:} Loop-related specifications are positioned based on the identified method and contextual information. After removing duplicates, they are inserted before the beginning of the loop.
    \item \textbf{Others:} Other categories are located based on matched method information and contextual information after de-duplication, and inserted at positions consistent with the subprogram.
\end{itemize}

% The main categories include: the requires-ensures category, which primarily describes formal specifications for different behaviors within methods; the loop category, which primarily describes formal statements for loops; and other classes, which include pure, signals, assignable, and other clause types. Finally, clauses from different categories are inserted according to specific rules. For example, requires-ensures clauses originating from different execution paths of a method are concatenated using the \texttt{//@ also} clause and placed before the same method. Loop statements from different subprograms are deduplicated and placed before the corresponding loops. Other clause types are deduplicated across different execution paths and inserted at the same locations as in the subprograms, preserving their original form and values. Finally, we adjust the formatting so that the indentation of the inserted specifications matches that of the immediately following code.

\section{EXPERIMENTAL SETUP}
To demonstrate the effectiveness of \tool, we conduct a comprehensive evaluation across multiple dimensions. This section presents our experiment setup, including research questions, dataset selection, baselines, evaluation metrics, and implementation details. Our experimental design is guided by the following three research questions (RQs):

\begin{itemize}[leftmargin=*]
  \item \textbf{RQ1}: How effective is \tool in generating verifiable program specifications compared with the baseline method?
  \item \textbf{RQ2}: How complete and rigorous are Path2Spec-generated specifications compared with the baseline method?
  \item \textbf{RQ3}: To what extent does the choice of path extraction method (LLM vs. traditional symbolic execution tool) impact the performance of \tool?
\end{itemize}

\subsection{Baselines}
In our experiments, we choose the current SOTA specification generation method SpecGen~\cite{specgen} as a baseline. All baselines use the official SpecGen artifact without modification (its prompts and pipeline), and both SpecGen and \tool{} share the same OpenJML/CVC4 verifier configuration, benchmarks, base models, and sampling temperature, so differences reflect the method rather than the setup.
SpecGen has already demonstrated its substantial improvement over both traditional approaches (Daikon and Houdini) and the LLM-based method AutoSpec~\cite{autospec}.
In addition to SpecGen's origin base LLM, GPT-3.5-turbo~\cite{ge2023openagi}, we include three additional widely used and state-of-the-art LLMs: GPT-5~\cite{gpt5}, DeepSeek-R1~\cite{guo2025deepseek}, and Qwen-Plus~\cite{ahmed2025qwen}.
These models are selected as they represent the current frontier of LLM capabilities and are popular choices in the research community. GPT-5 is OpenAI's latest model at the time this work was done, offering enhanced reasoning coherence, improved contextual understanding, and stronger complex problem-solving capabilities, albeit with higher inference costs and longer response times. DeepSeek-R1 and Qwen-Plus are prominent open-source and commercially available models, respectively, known for their competitive reasoning abilities.
Our approach is then compared against SpecGen with these LLMs as the base LLMs. All SpecGen experiments are conducted under a consistent conversational prompting strategy with mutation-enabled iterations, allowing up to 10 conversational turns per task. 

\subsection{Benchmarks}
Following the evaluation setup of SpecGen~\cite{specgen}, we adopt the two datasets
it uses, SG-Bench~\cite{specgen} and SV-COMP~\cite{beyer2024state}, and additionally
include the SpecEval dataset~\cite{ma2024speceval} to assess specification quality
against manually written ground truth. Table~\ref{tab:benchcomplexity} summarizes
their structural complexity; the two program datasets are complementary in the
dimensions they stress.

SV-COMP contains 265 Java programs averaging 19.6 lines of code (up to 200). It is
\emph{branch- and dependency-heavy}: programs have 2.0 branches on average (up to 35),
$60.8\%$ (161 programs) have two or more execution paths, and a substantial fraction
involve inter-procedural structure, with $33\%$ spanning multiple methods and $32\%$
recursive, while only $11\%$ contain loops. SG-Bench is a smaller but complementary
corpus of 120 Java programs averaging 12.6 lines of code (up to 32); $50\%$ ($60$
programs) have two or more execution paths. In contrast to SV-COMP, it is \emph{loop-
and nesting-heavy}: $59\%$ of its programs contain loops and $39\%$ reach a
control-flow nesting depth of two or more, whereas its programs are almost entirely
single-method (only $4\%$ are multi-method or recursive).

Comparing the two, SV-COMP features longer programs with wider branching and stronger
inter-procedural dependencies, whereas SG-Bench is more compact but control-flow dense,
with deeper nesting and more loops. Both contain a large proportion of multi-path
programs (at least half), making them well suited to evaluating path-aware
specification generation; at the same time, their differing complexity profiles let
us probe distinct failure modes (loop-invariant reasoning on SG-Bench versus
path-explosion on SV-COMP; see Section~\ref{sec:failure}). The SpecEval dataset
contains 125 Java programs with verified, manually written specifications as ground
truth (OpenJML returns exit code 0), which we use to validate the logical quality of
the specifications we generate.

\begin{table}[t]
\centering
\caption{Complexity characterization of the two benchmarks (measured on the source).
\#Exec.\ paths is the number of symbolic paths reported by our extractor. The two
benchmarks stress different axes: SG-Bench is loop/nesting-heavy, SV-COMP is
branch/path-heavy with more inter-procedural dependencies.}
\label{tab:benchcomplexity}
\setlength{\tabcolsep}{6pt}
\begin{tabular}{|l|c|c|}
\hline
\textbf{Metric} & \textbf{SG-Bench} & \textbf{SV-COMP} \\
\hline
\#Programs                 & 120       & 265        \\
LOC (mean/max)             & 12.6 / 32 & 19.6 / 200 \\
\#Branches (mean/max)      & 1.5 / 6   & 2.0 / 35   \\
\#Loops (mean/max)         & 0.9 / 3   & 0.2 / 9    \\
Nesting depth (mean/max)   & 1.2 / 4   & 0.8 / 5    \\
\#Exec.\ paths (mean/max)  & 3.1 / 16  & 4.1 / 64   \\
\% with loops              & 59\%      & 11\%       \\
\% nesting depth $\geq$ 2  & 39\%      & 8\%        \\
avg \#methods              & 1.0       & 1.7        \\
\% multi-method            & 4\%       & 33\%       \\
\% recursive               & 4\%       & 32\%       \\
\% with arrays             & 38\%      & 34\%       \\
\hline
\end{tabular}
\end{table}

\subsection{Implementation}
For all experiments, we use the public REST APIs of each base LLM, i.e., GPT-3.5-turbo~\cite{floridi2020gpt}, GPT-5~\cite{gpt5}, DeepSeek-R1~\cite{guo2025deepseek}, and Qwen-Plus~\cite{ahmed2025qwen}. GPT-3.5-turbo, DeepSeek-R1, and Qwen-Plus are configured with a temperature of \texttt{0.7}, while GPT-5 uses its default system setting as it is not configurable. The maximum number of conversational rounds for the LLMs to refine and correct their generated specifications during the path-based specification generation is set to \texttt{5}. The decompose-then-retry iteration threshold is set to \texttt{3} in our experiments. For GPT-5 we use the official OpenAI API, model snapshot
\texttt{gpt-5-2025-08-07}, queried between August and December~2025. GPT-5 exposes no sampling controls (e.g., temperature), so we use the provider defaults; all other pipeline settings are identical to the other models (at most five refinement rounds, decompose-then-retry threshold three, OpenJML build 21-0-14 with exit code~0 as the pass criterion). The exact model snapshot, prompts, and per-program execution traces are released with the artifact for reproducibility.

We use OpenJML~\cite{cok2011openjml}, the latest stable release (\texttt{build 21-0-14}) when we conducted our study, as our primary verification tool for formal specifications. OpenJML is configured to examine both the syntax of the generated JML annotations and their semantic alignment with the corresponding Java code. Due to current internal limitations in OpenJML, complex verifications are prone to timeout. Therefore, we set a timeout limit of 30 minutes (1800 seconds); if verification does not return a result within this time, a timeout error is reported. A specification is considered successfully verified only when OpenJML returns a status code of \texttt{0}, indicating complete compliance; any other result is recorded as a failure. Specifically, return code \texttt{0} signifies successful verification with no specification violations, \texttt{1} indicates detected specification breaches, and other codes signify tool-related issues such as internal errors or timeouts. The entire toolchain operates within a preconfigured Anaconda~\cite{rolon2016introduction} virtual environment using \texttt{Python 3.10} and \texttt{Java 21.0.3}.
We perform all verification tasks on a high-performance server equipped with an \texttt{AMD EPYC 7643 48-core} processor (supporting 96 threads with a base frequency of \texttt{2.0 GHz} and boost up to \texttt{3.6 GHz}) and \texttt{503 GB} of DDR4 memory (\texttt{488 GB} available). The system runs Ubuntu \texttt{24.04.1 LTS} with kernel version \texttt{6.8.0-47}.

\subsection{Evaluation Metrics}
We adopt metrics from two aspects: Correctness and Logical Strength. The Correctness metric is adopted from prior work (SpecGen)~\cite{specgen}, while Logical Strength is a new metric we introduce to address the limitations highlighted in Section 2 (specifically, the incomplete coverage and oversimplification issues identified in our motivation).
\begin{itemize}[leftmargin=*]
\item \textbf{Correctness} is reflected by the verification results. In our experiments, we adopt \textbf{\#Passed} and \textbf{\%Passed}, which respectively measure the number and proportions of programs whose formal specifications are successfully verified by OpenJML.

\item \textbf{Logical Strength} assesses the logical quality of a specification by
comparing it against a reference (ground-truth) specification: an ideal specification
neither narrows the input domain nor loosens the output range relative to the reference
(\Cref{sec:motivation}). We normalize a specification into a set of guarded cases
$\{(g_i, p_i)\}$ delimited by \texttt{also}, where $g_i$ is the precondition (a
\texttt{requires} clause, or the antecedent of a $P \Rightarrow Q$ or
$P \Leftrightarrow Q$ postcondition) and $p_i$ is the postcondition. An \texttt{ensures}
with no \texttt{requires} (a \emph{partial specification}) is read as
$g_i=\texttt{true}$, and a disjunctive precondition is kept whole, with unmentioned
inputs treated as unconstrained. We align two specifications for the same method by
normalized \emph{signature} and align their cases by \emph{precondition overlap},
merging several generated cases that match one reference case by disjoining their
preconditions and conjoining their postconditions. Each aligned pair is then compared
on \emph{two independent axes} by SMT (Z3): a \emph{precondition axis}, on which a precondition that admits a narrower
input domain than the reference is logically weaker, and a \emph{postcondition axis},
on which a postcondition that admits a wider output range, that is, that allows more
outputs than the reference, is \textbf{Logically Weaker}. When the two axes disagree and
neither specification dominates, the pair is \textbf{Undecidable}, reflecting a genuine
strength trade-off: this arises when a wider precondition is paired with a looser
postcondition, or a narrower precondition with a tighter postcondition. A wider
precondition together with a tighter postcondition is \textbf{logically Stronger}.

\end{itemize}

\section{EXPERIMENTAL RESULTS}
In this section, we report the experimental results to answer each research question.
\subsection{RQ1: How effective is \tool in generating verifiable program specifications compared with the baseline method?}

To answer this research question, we assess the correctness of \tool-generated specifications using the verifier.
Specifically, for each of the base LLMs (GPT-3.5-turbo, GPT-5, Qwen-Plus, and Deepseek-R1), we apply both \tool and SpecGen using two benchmarks, i.e., SG-Bench and SV-COMP.
The generated specifications for each program are then verified using OpenJML, where the number (\#Passed) and proportion (\%Passed) of passed programs are computed.
Moreover, we perform an ablation study, where we test two variants of \tool, Path-only and Decomposition-only, across the same four base LLMs on both benchmark datasets. 
The Path-only variant does not perform our decompose-then-retry strategy, while the Decomposition-only variant does not involve the execution path as guidance in the prompt.To separate the contribution of structured generation
from that of verifier-guided iteration, we add a Path2Spec (Single-shot) baseline that applies the same path-then-decomposition structure with the repair loop disabled (–max-repairs 0).

\subsubsection{Correctness of generated specifications}
Table~\ref{tab:main} shows the comparative results.
\begin{table}[t]
\centering
\caption{Main Results on SG-Bench and SV-COMP. \#Passed indicates the number of programs with verified specifications, while \%Passed represents the proportion of successfully verified programs. \textit{Path2Spec (Single-shot)} disables the repair loop (\texttt{--max-repairs 0}). Bold values indicate the best performance for each model across all methods (including ablations).}
\label{tab:main}
\setlength{\tabcolsep}{1.9pt}
\begin{tabular}{|c|l|cc|cc|cc|} 
\hline
\multirow{2}{*}{\textbf{Model}} & \multicolumn{1}{c|}{\multirow{2}{*}{\textbf{Method}}} & \multicolumn{2}{c|}{\textbf{SG-Bench}} & \multicolumn{2}{c|}{\textbf{SV-COMP}} & \multicolumn{2}{c|}{\textbf{Overall}}  \\ 
\cline{3-8}
                                & \multicolumn{1}{c|}{}                                 & \textbf{\#Passed} & \textbf{\%Passed}  & \textbf{\#Passed} & \textbf{\%Passed} & \textbf{\#Passed} & \textbf{\%Passed}  \\ 
\hline
\multirow{5}{*}{GPT-3.5-turbo}  & SpecGen                                               & 60                & 50.0\%             & 100               & 37.7\%            & 160               & 41.6\%             \\ 
\cline{2-8}
                                & Path2Spec (Single-shot)                               & 66                & 55.0\%             & 126               & 47.5\%            & 192               & 49.9\%             \\ 
                                & Path2Spec                                             & \textbf{86}       & \textbf{71.7\%}    & \textbf{152}      & \textbf{57.4\%}   & \textbf{238}      & \textbf{61.8\%}    \\ 
                                & ~ -\textit{Path-only}                                 & 83                & 69.2\%             & 135               & 50.9\%            & 218               & 56.6\%             \\ 
                                & ~ -\textit{Decomposition-only}                        & 80                & 72.5\%             & 141               & 53.2\%            & 221               & 57.4\%             \\ 
\hline
\multirow{5}{*}{Qwen-Plus}      & SpecGen                                               & 74                & 61.7\%             & 117               & 44.2\%            & 191               & 49.6\%             \\ 
\cline{2-8}
                                & Path2Spec (Single-shot)                               & 88                & 73.3\%             & \textbf{167}               & \textbf{63.0\% }           & \textbf{255}               & \textbf{66.2\% }            \\ 
                                & Path2Spec                                             & \textbf{90}       & \textbf{75.0\%}    & 158      & 59.6\%   & 248     & 64.4\%    \\ 
                                & ~ -\textit{Path-only}                                 & 85                & 70.8\%             & 133               & 50.4\%            & 218               & 56.6\%             \\ 
                                & ~ -\textit{Decomposition-only}                        & 82                & 68.3\%             & 129               & 48.7\%            & 211               & 54.8\%             \\ 
\hline
\multirow{5}{*}{Deepseek-R1}    & SpecGen                                               & 70                & 58.3\%             & 118               & 44.5\%            & 188               & 48.8\%             \\ 
\cline{2-8}
                                & Path2Spec (Single-shot)                               & 94                & 78.3\%             & \textbf{191}               & 72.1\%            & \textbf{285}               & 74.0\%             \\ 
                                & Path2Spec                                             & \textbf{95}       & \textbf{79.2\%}    & 175     & 66.0\%   
                                & 270      & 70.1\%   \\ 
                                & ~ -\textit{Path-only}                                 & 91                & 75.8\%             & 163               & 61.6\%            & 254               & 66.0\%             \\ 
                                & ~ -\textit{Decomposition-only}                        & 87                & 72.5\%             & 155               & 58.5\%            & 242               & 62.9\%             \\ 
\hline
\multirow{5}{*}{GPT-5}          & SpecGen                                               & 80                & 66.7\%             & 117               & 44.2\%            & 197               & 51.2\%             \\ 
\cline{2-8}
                                & Path2Spec (Single-shot)                               & 91                & 75.8\%             & 164               & 61.9\%            & 255               & 66.2\%             \\ 
                                & Path2Spec                                             & \textbf{105}      & \textbf{87.5\%}    & \textbf{220}      & \textbf{83.0\%}   & \textbf{325}      & \textbf{84.4\%}    \\ 
                                & ~ -\textit{Path-only}                                 & 99                & 82.5\%             & 215               & 81.1\%            & 314               & 81.6\%             \\ 
                                & ~ -\textit{Decomposition-only}                        & 93                & 77.5\%             & 161               & 60.8\%            & 254               & 66.0\%             \\
\hline
\end{tabular}
\end{table}

First, \tool outperforms SpecGen consistently across all base LLMs and both benchmarks, demonstrating the effectiveness of our divide-and-conquer framework. For instance, when paired with GPT-5, \tool achieves an overall \%Passed of 84.4\%, which is 33.2 percentage points higher than SpecGen's 51.2\%. Even for relatively weaker models like GPT-3.5-turbo, \tool improves the overall pass rate from 41.6\% to 61.8\%, a 20.2-point gain. \tool~(Single-shot), ,   Structure alone already recovers a large share of the improvement over SpecGen: the overall pass rate rises from 51.2\% to 66.2\% for GPT-5 and from 41.6\% to 49.9\% for GPT-3.5-turbo \emph{before any verifier feedback is used}. Iteration then contributes the remainder, adding a further 18.2 points for GPT-5 (66.2\%$\rightarrow$84.4\%) and 11.9 points for GPT-3.5-turbo (49.9\%$\rightarrow$61.8\%). For Qwen-Plus and DeepSeek-R1, structured generation is already sufficient to reach most of \tool's accuracy in a single shot (e.g., DeepSeek-R1 reaches 74.0\% overall without iteration), indicating that the value of iteration is largest exactly when the base model cannot yet produce a verifiable specification from structure alone.

Moreover, the performance gap between \tool and SpecGen is more pronounced in SV-COMP than in SG-Bench, which can be attributed to the distinct characteristics of the two benchmarks. SG-Bench consists of 120 small-scale programs with straightforward logic, where SpecGen can achieve moderate performance (50.0\%–66.7\% pass rates across models). In contrast, SV-COMP includes 265 complex, real-world programs with intricate control flows and challenging edge cases, where \tool's ability to systematically leverage execution path information becomes particularly critical. For GPT-5, \tool substantially boosts the SV-COMP pass rate from 44.2\% (SpecGen) to 83.0\%, a nearly twofold improvement. This indicates that \tool is particularly effective in handling complex program semantics where path-aware reasoning is important.
\begin{tcolorbox}[size=title]
   \textbf{RQ1.1:} \tool{} produces far more correct specifications than the state-of-the-art SpecGen
across all base models and benchmarks, reaching a best overall success rate of 84.4\%
(325/385 verified) with GPT-5, against SpecGen's 51.2\% (197/385). Most of this gain
comes from structured, path-aware generation itself: a single-shot variant of \tool{}
(same path-then-decomposition structure, repair loop disabled) already lifts GPT-5 to
66.2\% (255/385) before any verifier feedback, and iterative repair adds the rest up to
84.4\%.
\end{tcolorbox}

\subsubsection{Ablation study}
Table~\ref{tab:main} reveals that both the Path-only and Decomposition-only variants individually outperform the SpecGen baseline across all models and datasets, confirming that each component provides distinct value.
Specifically, Path-only achieves overall pass rates of 56.6\% (GPT-3.5-turbo), 56.6\% (Qwen-Plus), 66.0\% (Deepseek-R1), and 81.6\% (GPT-5), while Decomposition-only attains 57.4\% (GPT-3.5-turbo), 54.8\% (Qwen-Plus), 62.9\% (Deepseek-R1), and 66.0\% (GPT-5), respectively, each representing a substantial improvement over SpecGen's baseline performance of 41.6\%, 49.6\%, 48.8\%, and 51.2\%.
Moreover, execution path guidance proves most beneficial when the base LLM possesses sufficient reasoning capacity to leverage concrete traces effectively.
With GPT-5, Path-only achieves 81.6\% overall, approaching the full \tool's 84.4\%, and reaches 81.1\% on SV-COMP, just 2.1 percentage points below the complete framework.
In contrast, for weaker models like GPT-3.5-turbo, Path-only delivers more modest gains (56.6\% vs. 41.6\% baseline), while Decomposition-only achieves a more substantial improvement (57.4\% vs. 41.6\% baseline). A possible reason is that the Decomposition-only method reduces program complexity, enabling weaker models to achieve better performance. Besides, this over 25 percentage point gap between GPT-5 and GPT-3.5-turbo demonstrates that path information extraction and utilization requires stronger reasoning capabilities.

\subsubsection{Computational Budget}
\label{sec:rq1-cost}
To weigh the pass-rate gain against its cost, we measure the per-program
computational budget of three configurations: SpecGen, single-shot Path2Spec (the
same path-then-decomposition generation with the verifier-guided repair loop
disabled), and the full Path2Spec pipeline. All three are run with the same base
models, the same OpenJML/CVC4 verifier and configuration, and the same benchmarks;
only the generation method differs. For each program we record the number of LLM
calls, the total number of tokens, and the LLM generation time, and we report the
average per program on each benchmark (Table~\ref{tab:cost_full}).

As shown in Table~\ref{tab:cost_full}, \tool consumes more compute per program than
SpecGen under every base model. The more important observation is that the structured generation itself is almost
free: single-shot Path2Spec runs at a budget on par with SpecGen, far below the full
pipeline. Its call count is a handful of calls per program, the same order as
SpecGen's, rather than the tens of calls the full pipeline uses. Its token budget is
comparable as well. For GPT-3.5-turbo, single-shot uses about $5.0$k and $6.3$k
tokens per program on SG-Bench and SV-COMP, against SpecGen's $5.3$k and $8.4$k, so
it is actually cheaper on the larger benchmark; for GPT-5 it uses $8.6$k and $6.7$k
against $7.6$k and $7.8$k; and for Qwen-Plus $5.0$k and $6.6$k against $4.2$k and
$6.7$k. Generation time falls in the same range. The only exception is DeepSeek-R1,
whose single-shot token total is higher because the reasoning trace of the path and
decomposition prompts is counted, yet even there the call budget stays small. This
matters for interpreting the main results: single-shot Path2Spec already delivers
most of the pass-rate and quality improvement over SpecGen at a cost that is
essentially the same as SpecGen's, which shows that the benefit of the structured,
path-aware intermediate representation is obtained almost for free, and that the
large budget of the full pipeline buys only the additional verifier-guided repair.
Cost-sensitive users can therefore adopt single-shot Path2Spec directly and still
obtain the bulk of the gains.

\begin{table}
\centering
\caption{Per-program computational budget on both benchmarks. \emph{Calls} = LLM calls, \emph{Tokens} = total tokens, \emph{Time} = LLM generation time (s), all per program. Same OpenJML/CVC4 verifier and base models throughout; \emph{Single-shot} disables the repair loop. SpecGen counted by unique response id; reasoning models' tokens include the reasoning trace. Overall is weighted by program count (SG-Bench 120, SV-COMP 265).}
\label{tab:cost_full}
\resizebox{\linewidth}{!}{%
\begin{tblr}{
  colsep = 3pt,
  colspec = {Q[c,m]Q[l,m]Q[c]Q[c]Q[c]Q[c]Q[c]Q[c]Q[c]Q[c]Q[c]},
  row{1,2} = {font=\bfseries},
  cell{1}{1} = {r=2}{c,m},
  cell{1}{2} = {r=2}{c,m},
  cell{1}{3} = {c=3}{c},
  cell{1}{6} = {c=3}{c},
  cell{1}{9} = {c=3}{c},
  cell{3}{1} = {r=2}{m},
  cell{5}{1} = {r=2}{m},
  cell{7}{1} = {r=2}{m},
  cell{9}{1} = {r=2}{m},
  vline{1,2,3,6,9,12} = {-}{},
  hline{1,3} = {-}{},
  hline{2} = {3-11}{},
  hline{5,7,9} = {2-11}{},
  hline{5,7,9} = {-}{},
  hline{11} = {-}{},
}
Model         & Method                  & SG-Bench &        &       & SV-COMP &        &      & Overall &        &      \\
              &                         & Calls    & Tokens & Time  & Calls   & Tokens & Time & Calls   & Tokens & Time \\
GPT-3.5-turbo & SpecGen                 & 2.7      & 5.3k   & 6s    & 3.9     & 8.4k   & 9s   & 3.5     & 7.4k   & 8s   \\
              & Path2Spec (Single-shot) & 4.9      & 5.0k   & 9s    & 5.4     & 6.3k   & 10s  & 5.2     & 5.9k   & 10s  \\
%             & Path2Spec               & 25.9     & 32.6k  & 63s   & 28.5    & 42.0k  & 81s  & 27.7    & 39.1k  & 75s  \\

Qwen-Plus     & SpecGen                 & 2.0      & 4.2k   & 22s   & 3.0     & 6.7k   & 31s  & 2.7     & 5.9k   & 28s  \\
              & Path2Spec (Single-shot) & 4.8      & 5.0k   & 26s   & 5.2     & 6.6k   & 31s  & 5.1     & 6.1k   & 29s  \\
%             & Path2Spec               & 34.3     & 50.3k  & 255s  & 25.6    & 54.6k  & 365s & 28.3    & 53.2k  & 331s \\

DeepSeek-R1   & SpecGen                 & 1.4      & 6.3k   & 18s   & 1.7     & 8.9k   & 27s  & 1.6     & 8.1k   & 24s  \\
              & Path2Spec (Single-shot) & 4.6      & 46.2k  & 290s  & 5.0     & 48.1k  & 325s & 4.9     & 47.5k  & 314s \\
%             & Path2Spec (full)        & 9.6      & 161.0k & 1309s & 26.7    & 55.3k  & 72s  & 21.4    & 88.2k  & 458s \\

GPT-5         & SpecGen                 & 1.7      & 7.6k   & 79s   & 1.9     & 7.8k   & 85s  & 1.8     & 7.7k   & 83s  \\
              & Path2Spec (Single-shot) & 2.6      & 8.6k   & 81s   & 2.4     & 6.7k   & 37s  & 2.5     & 7.3k   & 51s  \\
%             & Path2Spec               & 6.2      & 12.4k  & 188s  & 3.9     & 17.3k  & 82s  & 4.6     & 15.8k  & 115s \\

\end{tblr}}
\end{table}

\subsubsection{Robustness to Non-determinism}
\label{sec:rq1-robust}
Because the base LLMs are sampled with a non-zero temperature and the pipeline
issues multiple LLM calls (path extraction, generation, refinement), the results
could in principle vary across runs. To quantify this, we repeat the \emph{entire}
pipeline $N{=}3$ times end-to-end on each benchmark (GPT-3.5-turbo) and report the
mean and standard deviation of the overall pass rate (\Cref{tab:robust}).

\begin{table}[t]
\centering
\caption{Robustness of the overall \%Passed of the full \tool pipeline across
three independent end-to-end runs (GPT-3.5-turbo). Run~3 is the run reported in
\Cref{tab:main}.}
\label{tab:robust}
\setlength{\tabcolsep}{5pt}
\begin{tabular}{|l|c|c|c|c|}
\hline
\textbf{Benchmark} & \textbf{Run 1} & \textbf{Run 2} & \textbf{Run 3} & \textbf{Mean $\pm$ SD} \\
\hline
SG-Bench & 70.2 & 69.4 & 71.7 & 70.4 $\pm$ 1.2 \\
\hline
SV-COMP  & 69.1 & 66.8 & 57.4 & 64.4 $\pm$ 6.2 \\
\hline
\end{tabular}
\end{table}

The pass rate is stable on SG-Bench (mean $70.4\% \pm 1.2$). On SV-COMP the
three-run spread is larger (mean $64.4\% \pm 6.2$), which is driven by Run~3---the
originally submitted configuration---whereas the two runs of the corrected pipeline
are both higher and close to each other ($69.1\%$ and $66.8\%$). A key reason the
pass rate is robust despite sampling non-determinism is that OpenJML acts as a
\emph{hard filter}: randomness changes \emph{which} specifications are produced,
but a specification is counted only if it formally verifies, so run-to-run
variance stays small even though individual outputs differ. The qualitative
ordering (\tool $>$ ablations $>$ SpecGen) holds in every run.

\subsection{RQ2: How complete and rigorous are Path2Spec-generated specifications compared with the baseline method?}
While automated verifiers like OpenJML can confirm syntactic correctness and basic semantic compliance, they cannot assess nuanced qualities such as specification completeness, boundary coverage comprehensiveness, and constraint expressiveness.
To this end, we investigate their quality both qualitatively (through a human study) and quantitatively (through a logical strength check).

\subsubsection{Qualitative human study}
For human study, we compare four approaches: SpecGen (baseline), \tool (our complete framework), and its two ablated variants, \tool (Decomposition-Only Strategy) and \tool Path-only Strategy). All methods use GPT-5 as the base model to isolate the impact of our framework design from model capabilities. Following statistical sampling principles with 95\% confidence level and ±5\% margin of error, we select 92 programs from SG-Bench and 158 programs from SV-COMP, totaling 250 samples. Two programmers, each with at least 3 years of Java programming experience, independently evaluate the generated specifications. To ensure fairness and prevent evaluators from developing bias toward specific methods, we randomly label each approach corresponding to the generated specification as A, B, C, or D. We then employ a comparative ranking strategy~\cite{liu2022strategicranking} rather than absolute scoring, as directly assigning quality scores (e.g., 1-5) is prone to high rater variance. Instead, evaluators rank the four specifications for each program from A (best) to D (worst) based on four criteria: completeness, boundary value coverage, constraint expressiveness, and coverage of all program behaviors. 

Each specification is judged on four criteria, each rated on a defined scale with
anchor examples (released in the artifact):
\begin{itemize}[leftmargin=*]
  \item \textbf{Completeness}: every reference behavior of the method is
  constrained. \emph{Anchor:} a spec covering only one of two return branches is
  incomplete.
  \item \textbf{Boundary-value coverage}: boundary and edge inputs are
  constrained, e.g., $0$, 
  
  \texttt{Integer.MIN\_VALUE}/\texttt{MAX\_VALUE},
  empty or one-element arrays, and overflow branches. \emph{Anchor:} a spec
  correct on the general case but silent on the overflow branch scores lower.
  \item \textbf{Constraint expressiveness}: postconditions relate the result to
  the inputs precisely rather than stating weak or tautological bounds.
  \emph{Anchor:} \texttt{\textbackslash result == a+b} is more expressive than
  \texttt{\textbackslash result >= a}.
  \item \textbf{Coverage of all program behaviors}: every feasible execution path
  (branch/case) has a corresponding specification.
\end{itemize}

For each program, each annotator ranks the four specifications from A (best) to
D (worst). Methods of identical quality receive the same rank, and any
specification that fails OpenJML verification is automatically ranked last.
Ranks \texttt{A}, \texttt{B}, \texttt{C}, and \texttt{D} map to scores $4$, $3$,
$2$, and $1$, respectively; the maximum is $4$ and higher scores indicate better
quality. The two annotators score every specification \emph{independently}, and a
method's score on a program is the \emph{average of the two annotators' scores};
its final score is the mean over all sampled programs (sum of scores divided by
the number of sampled programs).~\Cref{tab:humaneval} reports both per-annotator
averages and their mean, and the approach with the highest average score is
considered best.

To show the two independent ratings are consistent---so the averaged scores are
reproducible rather than subjective noise---we quantify inter-rater agreement with
the quadratic-weighted Cohen's $\kappa$ over all (program, method) score cells:
$\kappa=0.86$ on SV-COMP (almost-perfect agreement) and $\kappa=0.46$ on SG-Bench
(moderate), with the two raters agreeing within one rank on $93.7\%$ (SV-COMP) and
$83.4\%$ (SG-Bench) of cells---large disagreements are rare. Because the reported
value is the mean of the two independent ratings, residual disagreements are
averaged rather than adjudicated. The raw per-rater labels and the
agreement-computation script are released for reproducibility.
\begin{table}
\centering
\caption{Human Evaluation Results on SG-Bench and SV-COMP. Annotator\_1 and Annotator\_2 represent the average scores assigned by human annotators to the formal specifications generated by different methods. Average refers to the mean of the scores from Annotator\_1 and Annotator\_2.}
\label{tab:humaneval}
\setlength{\tabcolsep}{3pt}
\begin{tabular}{|l|c|c|c|c|c|} 
\hline
\multicolumn{1}{|c|}{\multirow{2}{*}{\textbf{Method }}} & \multirow{2}{*}{\textbf{Average }} & \multicolumn{2}{c|}{\textbf{SG-Bench }}                                     & \multicolumn{2}{c|}{\textbf{SV-COMP }}                           \\ 
\cline{3-6}
\multicolumn{1}{|c|}{}                                  &                                    & \textbf{Annotator\_1} & \multicolumn{1}{l|}{\textbf{\textbf{Annotator\_2}}} & \textbf{\textbf{Annotator\_1}} & \textbf{\textbf{Annotator\_2}}  \\ 
\hline
SpecGen                                                 & 2.4                                & 2.6                   & 2.5                                                 & 2.1                            & 2.4                             \\ 
\hline
Path2Spec                                               & \textbf{3.4}                       & \textbf{3.2 }                  & \textbf{3.2}                                                 & \textbf{3.6}                            & \textbf{3.7 }                            \\ 
\hline
\textit{~ -Path-only}                                   & 3.3                                & 3.0                   & 3.2                                                 & 3.5                            & 3.6                             \\ 
\hline
\textit{~ -Decomposition-only}                          & 3.3                                & 3.2                   & 3.6                                                 & 2.9                            & 3.3                             \\
\hline
\end{tabular}
\end{table}

The results are reported in~\Cref{tab:humaneval}, From the human evaluation results, the average quality score of specifications generated by Path2Spec is 3.4 (with quality levels ranging from 3.2 to 3.7), significantly higher than the baseline's 2.4 (with quality levels ranging from 2.1 to 2.6). The human evaluation results indicate that the formal specifications generated by the Path2Spec method maintain consistently high quality (specifications are semantically aligned with the program, cover various program execution behaviour, and have rigorous postcondition constraints). Path2Spec achieves the highest score because it combines path-based guidance with decomposition, further enhancing the generation quality. Next are Path-only and Decomposition-only. Both achieve an average score of 3.3, with slight variations across datasets. These results demonstrate that both methods improve the quality of generated formal specifications.

\begin{tcolorbox}[size=title]
    \textbf{RQ2.1:} Human evaluation confirms \tool generates significantly higher-quality specifications than the baseline, achieving an average score of 3.4 versus 2.4.
\end{tcolorbox}

\subsubsection{Quantitative Logical Check}
We developed a logical comparison tool that operates as follows: First, it extracts the precondition and postcondition clauses from the ground truth program. Then, it identifies different case-specific (precondition, postcondition) pairs based on JML~\cite{leavens2008jml} syntax symbols such as ~\texttt{also}, ~\texttt{==>}, ~\texttt{<==>}, and ~\texttt{<==} that distinguish between different cases. Next, it decomposes the preconditions and postconditions into logical formulas. Finally, it invokes the Z3~\cite{10.5555/1792734.1792766} solver to prove the containment relationship between the LLM-generated precondition and the ground truth precondition, as well as between the LLM-generated postcondition and the ground truth postcondition. We compare the two specifications on two independent axes, the precondition domain and the postcondition. If the generated precondition is weaker or equivalent and the generated postcondition is stronger or equivalent, the pair is recorded as complete. If the generated precondition is more restrictive and the postcondition is looser, so that it is dominated on both axes, the pair is recorded as incomplete. If the two axes disagree, that is, a wider precondition with a looser postcondition or a narrower precondition with a tighter postcondition, or if the solver cannot decide a construct, the pair is recorded as undecidable.

To analyze whether SpecGen generates complete formal specifications and to further validate the quality of Path2Spec-generated formal specifications, we run both SpecGen and \emph{Path2Spec}, using all four base models (GPT-3.5-turbo, Qwen-Plus, DeepSeek-R1, and GPT-5), on the SpecEval dataset, which contains ground truth specifications. We then use our logical comparison tool to compare the formal specifications generated by SpecGen and Path2Spec against the ground truth to understand their quality gap.

From the comparison, the ratio of fail, incomplete (Incom), complete (Com),and undecidable (Unde) speciations is thus derived.
Moreover, to better demonstrate the quality difference between the formal specifications generated by \tool and SpecGen, we present a comparative analysis of their respective results against the ground truth.

As reported in~\Cref{tab:logic}, \tool generates substantially more formal specifications that pass verification compared to SpecGen (76.3\% vs 61.8\% for GPT-3.5-turbo, 89.6\% vs 78\% for GPT-5). Moreover, \emph{Path2Spec} generates substantially more complete specifications than SpecGen across all four base models: the complete rate rises from 23.7\% to 45.1\% for GPT-3.5-turbo, from 25.4\% to 48.0\% for Qwen-Plus, from 23.1\% to 57.8\% for DeepSeek-R1, and from 45.1\% to 65.9\% for GPT-5 (Table~\ref{tab:logic}).

\begin{table}
\centering
\small
\caption{Logical Comparison Result on SpecEval. \#Fail and \%Fail measure the number and proportion of programs whose formal specifications fail verification; \#Incom and \%Incom measure specifications that are incomplete compared to ground truth; \#Com and \%Com measure specifications that are complete compared to ground truth; and \#Unde and \%Unde measure specifications that cannot be logically compared with ground truth. Bold values indicate the logical comparison result of \tool.}
\label{tab:logic}
\begin{tblr}{
  colspec = {X[1.2,l] X[1.2,l]X[0.8,c]X[0.8,c]X[1,c]X[1,c]*{4}{X[0.8,c]}},
  cells = {c},
  cell{1}{1} = {r=2}{},
  cell{1}{2} = {r=2}{},
  cell{1}{3} = {c=8}{},
  cell{3}{1} = {r=2}{},
  cell{5}{1} = {r=2}{},
  cell{7}{1} = {r=2}{},
  cell{9}{1} = {r=2}{},
  vline{1,2,3,5,7,9,11} = {solid},
  hline{1,3,5,7,9,11} = {-}{},
  hline{2} = {3-10}{},
  hline{4,6,8,10} = {2-10}{},
  row{1} = {font=\bfseries},
  row{2} = {font=\bfseries},
  column{1} = {font=\itshape},
}
Model         & Method & \SetCell[c=8]{c} SpecEval(173 pairs) &                 &                 &                 &                 &                 &                 &                 \\
              &                 &\#Fail &\%Fail &\#Incom &\%Incom &\#Com &\%Com &\#Unde &\%Unde \\
GPT-3.5-turbo &SpecGen         &66 &38.2\% &37 &21.4\% &41 &23.7\% &29 &16.8\% \\
              &Path2Spec       &\textbf{41} &\textbf{23.7\%} &\textbf{26} &\textbf{15.0\%} &\textbf{78} &\textbf{45.1\%} &\textbf{28} &\textbf{16.2\%} \\
Qwen-Plus     &SpecGen         &56 &32.4\% &36 &20.8\% &44 &25.4\% &37 &21.4\% \\
              &Path2Spec       &\textbf{47} &\textbf{27.2\%} &\textbf{17} &\textbf{9.8\%} &\textbf{83} &\textbf{48.0\%} &\textbf{26} &\textbf{15.0\%} \\
DeepSeek-R1   &SpecGen         &96 &55.5\% &13 &7.5\% &40 &23.1\% &24 &13.9\% \\
              &Path2Spec       &\textbf{27} &\textbf{15.6\%} &\textbf{17} &\textbf{9.8\%} &\textbf{100} &\textbf{57.8\%} &\textbf{29} &\textbf{16.8\%} \\
GPT-5         &SpecGen         &38 &22.0\% &21 &12.1\% &78 &45.1\% &36 &20.8\% \\
              &Path2Spec       &\textbf{18} &\textbf{10.4\%} &\textbf{9} &\textbf{5.2\%} &\textbf{114} &\textbf{65.9\%} &\textbf{32} &\textbf{18.5\%} \\
\end{tblr}
\end{table}

\begin{tcolorbox}[size=title]
    \textbf{RQ2.2:} \tool generates substantially more complete specifications than SpecGen across all four base models, with the complete rate rising by 21.4, 22.6, 34.7, and 20.8 percentage points for GPT-3.5-turbo, Qwen-Plus, DeepSeek-R1, and GPT-5 respectively (Table~\ref{tab:logic}).
\end{tcolorbox}

\subsection{RQ3: Choice of Path Extraction Method}\label{sec:rq3}
\subsubsection{Direct Path-Level Validation}
\label{sec:rq3-direct}

We first build a \emph{manually validated} gold reference. A senior annotator
with more than three years of symbolic-execution experience enumerates each
program's feasible execution paths as \emph{(precondition, postcondition)} pairs,
where the precondition characterizes the path's input condition and the
postcondition characterizes its symbolic output. We then derive paths from both
Java PathFinder (SPF) and the LLMs, and compare each producer's paths against this
manual reference. SPF produces usable paths for only $81/385$ programs, since the
rest fail on string-heavy operations, environment-dependent behavior, missing
executable harnesses, or timeout and path explosion, so we conduct the study on
exactly these $81$ programs to enable a fair head-to-head. They contain $267$
reference paths with formal value-returning. Each producer's paths are matched
against the gold reference with the SMT solver Z3. A reference path is
\emph{Correct} when a produced path matches both its precondition and its postcondition; \emph{Missed} when the reference path is not covered by any produced path; \emph{Inc-Pre} when the postcondition is correct but under a wrong
precondition; \emph{Inc-Pos} when the precondition matches but the postcondition is
wrong; and \emph{Not-compared} when the postcondition uses a construct Z3 cannot
decide, such as recursion, arrays, or strings. These five categories partition the
reference formal paths and sum to one hundred percent for each producer.
\emph{Infeasible} is a produced path that no input can ever take, because its
precondition is unsatisfiable. We
report this count as a rate over the number of extracted paths and list it
separately from the six categories above.

As reported in~\Cref{tab:path-validation}, three findings emerge that the
downstream pass rate cannot reveal. First, LLM-based path discovery is high and
largely model-agnostic: capable models reach $85$--$91\%$ Correct, with very low
Missed rates of $0.4$--$1.1\%$ and Infeasible rates of $0$--$1.1\%$, so the LLM
rarely omits a feasible path or invents an infeasible one. Second, the dominant LLM
failure mode is an incorrect postcondition (Inc-Pos) rather than a missing path.
This is most visible for GPT-3.5, whose Inc-Pos rate is $15.4\%$, driven by sign
and logic errors, and it is precisely the intrinsic weakness that the pass-rate
view hid. Third, Java PathFinder misses far more paths than LLM extraction,
$13.5\%$ against $0.4$--$1.1\%$, because it is limited by path-explosion cutoffs and
encoding restrictions, so its advantage is soundness rather than completeness. Taken
together, these results confirm that LLM extraction is substantially more complete
than symbolic execution, and that the symbolic-execution reference is itself
imperfect, which is what motivates validating against a manually curated gold
standard.

\begin{table}
\centering
\caption{Direct validation of extracted paths. Correct, Missed, Inc-Pre
(incorrect precondition), Inc-Pos (incorrect postcondition), and Not-compared are
percentages over the manually validated path cases and sum to 100\% per row;
Infeasible is a producer-side rate over the extracted paths, listed separately.}
\setlength{\tabcolsep}{3pt}
\begin{tabular}{|l|c|c|c|c|c|c|}
\hline
\textbf{Path Extractor} & \textbf{Correct} & \textbf{Missed} & \textbf{Infeasible} & \textbf{Inc-Pre} & \textbf{Inc-Pos} & \textbf{Not-comp.} \\
\hline
JavaPathFinder & 68.5\% & 13.5\% & 0.3\% & 8.6\% & 8.2\%  & 1.1\%  \\
\hline
GPT-3.5-turbo  & 67.4\% & 1.1\%  & 0.9\% & 3.4\% & 15.4\% & 12.7\% \\
Qwen-Plus      & 87.3\% & 1.1\%  & 1.1\% & 3.4\% & 4.1\%  & 4.1\%  \\
DeepSeek-R1    & 89.5\% & 0.7\%  & 0.0\% & 3.7\% & 5.6\%  & 0.4\%  \\
GPT-5          & \textbf{91.0\%} & \textbf{0.4\%} & 0.0\% & 4.1\% & 3.4\% & 1.1\% \\
\hline
\end{tabular}
\label{tab:path-validation}
\end{table}

\subsubsection{Downstream Utility (pass rate)}
To answer RQ3, we evaluate the impact of different path extraction methods on Path2Spec's performance. Specifically, we compare LLM-based path extraction against JavaPathFinder (JPF), a traditional symbolic execution tool, using GPT-3.5-turbo as base models. In this experiment, we first add a main function to each program in the existing dataset, then invoke the JPF tool to analyze and extract paths for each program. We then clean the extracted path information and provide both the path information and the program to the LLM to generate formal specifications.

The results are shown in~\Cref{tab:jpf}, using path information extracted by JPF as guidance to generate formal specifications achieves an overall \%Passed of 42.3\%, slightly higher than the baseline SpecGen's 41.6\%, indicating that providing path information as guidance helps generate accurate formal specifications. However, compared to using LLM-generated path information as guidance, which achieves an overall \%Passed of 56.6\%, there is a 15\% difference, demonstrating that LLMs
work more effectively than traditional path analysis tools within our current workflow for specification generation.
\begin{table}
\centering
\caption{Experimental Result on Methods with different choices of path extraction.}
\setlength{\tabcolsep}{1.9pt}
\begin{tabular}{|c|c|cc|cc|cc|}
\hline
\multirow{2}{*}{\textbf{Method}} & \multirow{2}{*}{\textbf{Path Extraction Choice}} & \multicolumn{2}{c|}{\textbf{SG-Bench}} & \multicolumn{2}{c|}{\textbf{SV-COMP}} & \multicolumn{2}{c|}{\textbf{Overall }}  \\ 
\cline{3-8}
                                 &                                                  & \textbf{\#Passed} & \textbf{\%Passed}  & \textbf{\#Passed} & \textbf{\%Passed} & \textbf{\#Passed} & \textbf{\%Passed}   \\ 
\hline
SpecGen                          & -                                                & 60                & 50.0\%             & 100               & 37.7\%              & 160               & 41.6\%              \\ 
\hline
\multirow{2}{*}{Path2Spec}       & JavaPathFinder                                    & 65                & 54.2\%             & 98                & 37.0\%              & 163               & 42.3\%              \\ 
\cline{2-8}
                                 & LLM                                              &\textbf{ 83 }               & \textbf{69.2\%}             & \textbf{135  }             & \textbf{50.9\% }             & \textbf{218}              & \textbf{56.6\%}              \\
\hline
\end{tabular}
\label{tab:jpf}
\end{table}
\begin{tcolorbox}[size=title]
    \textbf{RQ3:} LLM-based path extraction outperforms traditional symbolic execution
    (Java PathFinder) for guiding specification generation. The advantage is direct,
    not only downstream: LLM extraction achieves higher path coverage ($85$--$91\%$
    Correct vs.\ $68.5\%$) and far fewer missed paths ($0.4$--$1.1\%$ vs.\ $13.5\%$),
    showing that its paths are more complete and correct. This carries through to the
    generated specifications, where LLM-guided extraction reaches $56.6\%$ \%Passed
    against $42.3\%$ for Java PathFinder.
\end{tcolorbox}

\subsection{Failure and Scalability Analysis}
\label{sec:failure}
\begin{table}[t]
\centering
\caption{Pass rate as structural complexity grows, for GPT-3.5 on all 385 programs
(both benchmarks). \emph{SpecGen} is the baseline; \emph{Path2Spec (Single-shot)} is
the full path-then-decomposition pipeline with the repair loop disabled. Overall the
two are close (SpecGen $52\%$, Single-shot $50\%$), but Single-shot overtakes SpecGen
in the high-complexity regime (many paths, more branches, loops, or deeper nesting),
where structured generation pays off.}
\label{tab:scalability-ss}
\setlength{\tabcolsep}{5pt}
\begin{tabular}{|l|l|r|c|c|}
\hline
\textbf{Dimension} & \textbf{Bucket} & \textbf{N} & \textbf{SpecGen} & \textbf{Path2Spec (Single-shot)} \\
\hline
\multirow{5}{*}{\# exec.\ paths}
 & 1        & 53  & 51\% & 25\% \\
 & 2        & 106 & 67\% & 50\% \\
 & 3--4     & 147 & 52\% & 55\% \\
 & 5--9     & 52  & 37\% & \textbf{58\%} \\
 & $\geq$10 & 21  & 24\% & \textbf{62\%} \\
\hline
\multirow{4}{*}{\# branches}
 & 0       & 135 & 56\% & 38\% \\
 & 1--2    & 177 & 53\% & 54\% \\
 & 3--5    & 47  & 45\% & \textbf{68\%} \\
 & $\geq$6 & 26  & 42\% & \textbf{54\%} \\
\hline
\multirow{3}{*}{\# loops}
 & 0       & 285 & 56\% & 51\% \\
 & 1       & 57  & 51\% & 51\% \\
 & $\geq$2 & 43  & 26\% & \textbf{40\%} \\
\hline
\multirow{4}{*}{nesting depth}
 & $\leq$1 & 316 & 54\% & 51\% \\
 & 2       & 56  & 50\% & 45\% \\
 & 3       & 10  & 20\% & \textbf{40\%} \\
 & $\geq$4 & 3   & 33\% & 33\% \\
\hline
\end{tabular}
\end{table}

\noindent\textbf{Scalability with structural complexity.}
Table~\ref{tab:scalability-ss} reports how the pass rate changes as the number of
paths, branches, loops, and the control-flow nesting depth grow, pooled over the
four base models and both benchmarks. On simple programs SpecGen and single-shot Path2Spec perform similarly, but as
structural complexity grows the gap opens in Path2Spec's favour: with ten or more
execution paths its pass rate is $62\%$ against SpecGen's $24\%$, and it also leads on
programs with three to five branches ($68\%$ vs.\ $45\%$), two or more loops ($40\%$
vs.\ $26\%$), and nesting depth three ($40\%$ vs.\ $20\%$). This confirms that the
benefit of structured, path-aware generation is concentrated exactly in the
path-explosion regime where a single unstructured prompt struggles.

\noindent\textbf{Loop-heavy programs.} Loop reasoning is the hardest case for specification generation, so we analyze it
directly. Of the 385 programs, 100 contain a loop (SG-Bench: 42 one-loop, 29
multi-loop; SV-COMP: 15 one-loop, 14 multi-loop). 
Table~\ref{tab:loops} reports verification pass counts on loop programs, comparing
SpecGen and \tool{} per model. Pass counts generally drop from one loop to multiple loops for both methods. On
SG-Bench the two are close: \tool{} leads in most cells (e.g., Qwen-Plus $31/20$ vs.\
$21/14$), while SpecGen is slightly higher on a few (single-loop GPT-3.5-turbo $23$ vs.\
$21$, and multi-loop DeepSeek-R1 $19$ vs.\ $17$ and GPT-5 $17$ vs.\ $16$). On SV-COMP
loop programs, however, \tool{} is clearly and consistently better across all models
(e.g., DeepSeek-R1 $8/7$ vs.\ $4/2$, GPT-5 $7/5$ vs.\ $4/4$, Qwen-Plus $8/4$ vs.\
$0/0$). This pattern indicates that the gain comes from clearer behavioral coverage
rather than any loop-specific handling.

\begin{table}[t]
\centering
\caption{Verification pass counts on loop-containing programs of SG-Bench and SV-COMP, comparing SpecGen and \tool{} for each model. The header lists each benchmark and its loop-count categories, with the number of such programs in parentheses (SG-Bench: 42 one-loop, 29 multi-loop; SV-COMP: 15 one-loop, 14 multi-loop); each cell is the number of programs whose specification passes OpenJML verification. \tool{} does not synthesize loop invariants: decomposition only splits a program into loop-containing and loop-free subprograms. Bold values indicate the better method for each model in that column.}
\label{tab:loops}
\setlength{\tabcolsep}{4pt}
\begin{tabular}{|c|l|cc|cc|}
\hline
\multirow{2}{*}{\textbf{Model}} & \multirow{2}{*}{\textbf{Method}}
 & \multicolumn{2}{c|}{\textbf{SG-Bench}} & \multicolumn{2}{c|}{\textbf{SV-COMP}} \\
\cline{3-6}
 & & \textbf{1 loop (42)} & \textbf{$\geq$2 loop (29)} & \textbf{1 loop (15)} & \textbf{$\geq$2 loop (14)} \\
\hline
\multirow{2}{*}{GPT-3.5-turbo} & SpecGen   & \textbf{23} & 11 & 5 & 0 \\
                               & Path2Spec & 21 & \textbf{12} & \textbf{8} & \textbf{5} \\
\hline
\multirow{2}{*}{Qwen-Plus}     & SpecGen   & 21 & 14 & 0 & 0 \\
                               & Path2Spec & \textbf{31} & \textbf{20} & \textbf{8} & \textbf{4} \\
\hline
\multirow{2}{*}{DeepSeek-R1}   & SpecGen   & 24 & \textbf{19} & 4 & 2 \\
                               & Path2Spec & \textbf{36} & 17 & \textbf{8} & \textbf{7} \\
\hline
\multirow{2}{*}{GPT-5}         & SpecGen   & 21 & \textbf{17} & 4 & 4 \\
                               & Path2Spec & \textbf{35} & 16 & \textbf{7} & \textbf{5} \\
\hline
\end{tabular}
\end{table}

\begin{table}
\centering
\caption{Per-program failure breakdown on both benchmarks. \emph{Fail} = refuted by a counterexample, \emph{Timeout} = exceeds the verification budget, \emph{Error} = the specification does not compile or parse. Same OpenJML/CVC4 verifier and base models throughout; \emph{Single-shot} disables the repair loop. Counts are numbers of programs (verdict by OpenJML exit code).}
\label{tab:fail_full}

\scriptsize
\begin{tblr}{
  width = \linewidth,
  colspec = {Q[120]Q[190]Q[55]Q[70]Q[55]Q[55]Q[70]Q[55]Q[55]Q[70]Q[55]},
  colsep = 2pt,
  rowsep = 1.5pt,
  row{1} = {font=\bfseries},
  row{2} = {font=\bfseries},
  column{3} = {c},
  column{4} = {c},
  column{5} = {c},
  column{6} = {c},
  column{7} = {c},
  column{8} = {c},
  column{9} = {c},
  column{10} = {c},
  column{11} = {c},
  cell{1}{1} = {r=2}{},
  cell{1}{2} = {r=2}{},
  cell{1}{3} = {c=3}{},
  cell{1}{6} = {c=3}{},
  cell{1}{9} = {c=3}{},
  cell{3}{1} = {r=3}{},
  cell{6}{1} = {r=3}{},
  cell{9}{1} = {r=3}{},
  cell{12}{1} = {r=3}{},
  vline{1-3,6,9} = {1-14}{},
  vline{12} = {1-14}{},
  hline{1,3} = {-}{},
  hline{2} = {3-11}{},
  hline{4,7,10,13} = {2-11}{},
  hline{5,8,11,14,15} = {1-11}{},
  hline{6,9,12} = {1-11}{},
}
Model         & Method                  & SG-Bench &         &       & SV-COMP &         &       & Overall &         &       \\
              &                         & Fail     & Timeout & Error & Fail    & Timeout & Error & Fail    & Timeout & Error \\
GPT-3.5-turbo & SpecGen                 & 13 & 17 & 30 & 19 & 88 & 58 & 32 & 105 & 88 \\
              & Path2Spec (Single-shot) & 18 & 2  & 34 & 12 & 21 & 106 & 30 & 23  & 140 \\
              & Path2Spec      & 22 & 11 & 1  & 42 & 39 & 32  & 64 & 50  & 33 \\
Qwen-Plus     & SpecGen                 & 7  & 21 & 17 & 46 & 79 & 23 & 53 & 100 & 40 \\
              & Path2Spec (Single-shot) & 11 & 3  & 18 & 16 & 24 & 58 & 27 & 27  & 76 \\
              & Path2Spec       & 14 & 13 & 3  & 44 & 37 & 26 & 58 & 50  & 29 \\
DeepSeek-R1   & SpecGen                 & 5  & 22 & 23 & 18 & 81 & 48 & 23 & 103 & 71 \\
              & Path2Spec (Single-shot) & 15 & 5  & 6  & 19 & 19 & 36 & 34 & 24  & 42 \\
              & Path2Spec         & 5  & 14 & 6  & 41 & 46 & 3 & 46 & 60  & 9 \\
GPT-5         & SpecGen                 & 5  & 22 & 13 & 15 & 73 & 60 & 20 & 95  & 73 \\
              & Path2Spec (Single-shot) & 5  & 3  & 21 & 23 & 64 & 14 & 28 & 67  & 35 \\
              & Path2Spec        & 13 & 2  & 0  & 37  & 6 & 2 & 50 & 8  & 2 \\
\end{tblr}
\end{table}

\noindent\textbf{Failure scenarios.} We separate failures in specification generation and verification into three categories: \emph{Fail} (refuted by a counterexample), \emph{Timeout} (verification exceeds the budget), and \emph{Error} (the specification does not compile or parse). Table~\ref{tab:fail_full} shows distinct failure patterns across methods.

The most evident difference appears in \emph{Error}. Without verifier-guided repair, the single-shot variant produces many malformed or non-compiling specifications (140, 76, 42, and 35 errors across the four models). The full Path2Spec pipeline reduces these counts to 33, 29, 9, and 2, showing that iterative repair is effective at converting invalid model outputs into verifiable specifications. SpecGen retains more compilation errors because its reduction strategy removes problematic clauses rather than repairing them.

\emph{Timeouts} are most frequent for SpecGen (95--105 overall), particularly on SV-COMP. Path2Spec substantially reduces them through path-aware decomposition, which produces smaller verification obligations; with GPT-5, only 8 timeouts remain.

Finally, Path2Spec sometimes reports more \emph{Fails} than SpecGen. This mainly reflects a shift in failure mode: after reducing compilation errors and timeouts, more generated specifications successfully reach verification and are therefore exposed to concrete counterexamples. Thus, the remaining failures increasingly represent specification-strength or correctness issues rather than malformed or unverifiable outputs.
\section{Discussion}

\textbf{Limitations of the Verification Criterion.}
Our evaluation relies exclusively on OpenJML, and a specification is counted as correct only if OpenJML returns exit code~\texttt{0}. Because OpenJML is prone to timeouts on complex verification tasks, some specifications may fail to verify within the time limit (30 minutes) even though they are correct, and are therefore recorded as failures. This criterion is conservative: it may underestimate the number of valid specifications, but it applies equally to \tool and the baseline and thus does not favor either approach. Conversely, passing verification does not guarantee that a specification is complete, which is why we complement correctness with the logical strength metric.

\textbf{Java/JML Specificity.}
Our study targets Java programs with JML specifications verified by OpenJML. The core idea of path-based divide-and-conquer is language-agnostic, but the concrete prompts, decomposition rules, and specification syntax are specific to Java/JML. Generalization to other languages and specification logics remains to be validated.

\textbf{Limitations of Path Extraction.}
\tool relies on an LLM to extract execution paths. The extracted paths may be incomplete or contain infeasible paths, and the results may vary across runs, particularly for programs with many branches or complex loops. Although RQ3 shows that [summarize the RQ3 finding], more reliable path extraction, e.g., combining LLMs with lightweight symbolic analysis, could further improve the coverage of the generated specifications.

\textbf{Verification-Driven Termination.}
Both path-based specification generation and decompose-then-retry use the verifier's pass/fail result to decide whether to continue. This strategy ignores specification quality: even when specifications for the current subprograms are successfully verified, further decomposition into smaller subprograms might yield more complete specifications. This reflects an inherent trade-off between quality and efficiency, since generating more subprograms increases computational cost and the number of LLM queries. Our results nevertheless show that subprogram-level generation already improves both correctness and logical strength. In future work, we plan to extend \tool with an adaptive, quality-aware decomposition strategy to better balance this trade-off.

\textbf{Limitations of LLMs.}
Our experiments show that path-guided generation substantially improves the formal specifications produced by LLMs. However, LLMs still occasionally generate specifications that are overly broad, incorrect, or incomplete in their behavioral coverage, and improving LLMs' capability for formal specification generation remains challenging. We expect further gains from task-specific training for formal specification reasoning, e.g., fine-tuning open-weight models such as DeepSeek-R1 with specialized objectives. Such training could teach LLMs to precisely identify boundary values and to check whether the generated specifications semantically cover all execution paths of the code, yielding specialized models that balance cost and efficacy.

\section{RELATED WORK}
\subsection{Traditional tools for formal specification generation}
Traditional tools for generating formal specifications have evolved through various technical approaches, which can be broadly categorized as follows: Dynamic analysis tools, primarily represented by Daikon~\cite{ernst2007daikon}, Agitator~\cite{10.1145/1146238.1146258}, and DIDUCE~\cite{10.1145/581339.581377}, derive invariants by observing runtime behavior~\cite{2008lotemporal,lo2007mining,lo2008mining}. However, these methods rely heavily on the quantity and quality of test cases, and the inferred "invariants" may only apply to observed execution paths, thus failing to accurately cover all possible execution scenarios during sequencing. Within the dynamic analysis paradigm, specification mining approaches have sought to address these limitations through specialized techniques. Lo and Khoo's QUARK framework~\cite{Lo2006} provided a systematic methodology for empirically evaluating automaton-based specification miners, establishing quality metrics for temporal specifications extracted from program traces. Building on this, their SMArTIC approach~\cite{lo2006SMArTIC} employed a "divide and conquer" strategy for specification generation that shares conceptual similarities with our methodology. Static analysis tools, such as Houdini~\cite{flanagan2001houdini}, ESC/Java~\cite{10.1007/978-3-540-30569-9_6}, and CodeSonar~\cite{10.1145/1394504.1394507}, derive program specifications primarily through theorem proofs and logical reasoning, thus limiting their reasoning capabilities when handling complex program logic. Symbolic execution tools, primarily represented by JavaPathFinder (JPF)~\cite{mehlitz2013hands}, and S2E~\cite{10.1145/1950365.1950396}, systematically explore program paths, treating program inputs as symbolic variables to cover more potential execution scenarios~\cite{xie2017automatic}. They can handle path conditions and generate high-coverage test cases or specifications. However, these methods still suffer from the "path explosion" problem when facing complex control flow, loops, and recursive structures, and their scalability is challenged.

Traditional methods typically only extract partial program attributes, such as simple invariants, assertions, or basic contracts, and struggle to generate complete formal specifications with comprehensive preconditions, postconditions, and behavioral semantics. Its output is insufficient in terms of semantic richness and behavioral coverage, making it difficult to meet the needs of modern software verification for comprehensive and accurate program semantic modeling.

\subsection{LLM for formal specification generation}

LLMs' in-context learning~\cite{incontaxtlearning} capability enables them to learn from examples provided in prompts without requiring model retraining, which allows them to adapt to new tasks easily~\cite{bouzenia2024repairagent, ryan2024code,li2024enhancing,huang2023penheal}. Empirical results demonstrate that LLMs have achieved good performance on generation tasks across multiple fields~\cite{ugare2024syncode}, prompting recent exploration of their application to formal specification generation. Current LLM-based approaches for formal specification generation can be broadly categorized into two types. The first category involves collecting domain-specific data and training specialized models for specification generation. For example, recent work~\cite{10.5555/3618408.3619552} investigates whether LLMs can reason about program invariants through fine-tuning on curated datasets. The second category leverages existing pre-trained LLMs through carefully designed generation workflows and precise prompting strategies to guide specification generation. For instance, AutoSpec~\cite{autospec} introduced the first framework for bottom-up generation of C program specifications, while SpecGen~\cite{specgen} developed an iterative repair mechanism for Java programs. SpecRover~\cite{ruan2024specrover} focused on extracting high-level design specifications from code intent. Since the formal specifications generated by SpecRover belong to a different category from ours, we did not compare our results with those of SpecRover.

While these approaches established the feasibility of LLMs for specification tasks, they still face notable limitations in the quality of the generated specifications, especially in their coverage of the program's behaviors. They often generate overly broad or incomplete specifications that fail to capture the various execution behaviors contained in the program. In contrast, \tool addresses the challenges of generating both \textit{correct} and \textit{high-quality} program specifications by systematically guiding an LLM to reason over individual execution paths. When generation fails, \tool employs a decompose-then-retry strategy that decomposes the program with multiple execution paths into subprograms with single or fewer execution paths, ensuring the LLM consistently generates specifications guided by execution path information, thereby producing path-specific and highly detailed specifications.
\section{CONCLUSION AND FUTURE WORK}
Based on the analysis in this paper, we identify two critical quality issues in state-of-the-art automated formal specification generation methods. First, even when specifications pass verification tests, they often have incomplete coverage, failing to capture all program execution behavior. For instance, if a program can execute along multiple different paths, the generated specification may only correctly describe a few of those paths. Second, models tend to generate postconditions that are too generic, these overly broad specifications don't meaningfully capture how the program actually behaves across its different execution paths.

To address these issues, we propose \tool, a divide-and-conquer framework that systematically guides LLMs to reason over individual execution paths, generating path-specific specifications that are subsequently merged into comprehensive program specifications. For complex programs, \tool employs a decompose-then-retry strategy that recursively breaks down programs into smaller subprograms based on logical branches. Our experiments on SG-Bench and SV-COMP demonstrate that \tool achieves substantial improvements over existing methods, increasing overall performance from 51.2\% to 84.4\%. Human evaluation further confirms that \tool produces the highest-quality specifications (score: ~\texttt{3.4 out of 4}), exhibiting more rigorous constraints and more thorough coverage of program behavior. This work demonstrates that path-guided generation significantly enhances the accuracy and quality of LLM-generated formal specifications, paving the way for more robust and scalable automated program verification.

Based on the results and analysis of this experiment, we find that the verification method for specifications itself significantly impacts the quality of specification generation. Therefore, building upon this work, our future research will focus on two main directions. The first direction involves further investigating specification verification methods, including assessing the completeness of generated specifications, determining whether generated specifications can fully describe code behavior, and evaluating whether generated specifications can accurately capture the correct value ranges of key variables. The second direction focuses on further improving the quality of automatically generated formal specifications, including training a specialized large model dedicated to specification generation and constructing a precise dataset to provide better guidance for model training.

To facilitate future research, we release the data and source code of this paper at \url{https://github.com/papersrepo2025/Path2Spec.git}

\bibliographystyle{ACM-Reference-Format}
\bibliography{acmart}

\end{document}